\documentclass[traditabstract,bibyear]{aa}
\usepackage{times}
\usepackage{longtable}
\usepackage{natbib}
\usepackage{rotate}
\usepackage{lscape}
\usepackage{amssymb}
\usepackage{graphicx}
\usepackage[draft]{hyperref}

\bibpunct{(}{)}{;}{a}{}{,} 

\def\5{\footnotesize V\normalsize}
\def\4{\footnotesize IV\normalsize}
\def\3{\footnotesize III\normalsize}
\def\2{\footnotesize II\normalsize}
\def\1{\footnotesize I\normalsize}

\def\lam{$\lambda$}
\def\kms{$\mbox{km s}^{-1}$}

\def\pp{$\phantom{-}$}
\def\o{$\phantom{1}$}

\begin{document}

\title{2dF-AAOmega spectroscopy of massive stars in the\\Magellanic
  Clouds} \subtitle{The north-eastern region of the Large Magellanic
  Cloud\thanks{Copies of the spectra are available at the CDS via
    anonymous ftp to cdsarc.u-strasbg.fr (130.79.128.5) or via
    http://cdsweb.u-strasbg.fr/cgi-bin/qcat?J/A+A/}}

\author{C.~J.~Evans\inst{1}, J.~Th.~van~Loon\inst{2}, R.~Hainich\inst{3} \& M. Bailey\inst{4,2}}
\offprints{chris.evans@stfc.ac.uk}
\authorrunning{C.~J.~Evans et al.}
\titlerunning{AA$\Omega$ spectroscopy of massive stars in the NE of the LMC}

\institute{UK Astronomy Technology Centre, Royal Observatory, 
           Blackford Hill, Edinburgh, EH9 3HJ,~UK
             \and
           Astrophysics Group, School of Physical \& Geographical Sciences, 
           Lennard-Jones Laboratories, Keele University, ST5 5BG, UK
             \and
           Institute for Physics and Astronomy, University of Potsdam, 14476 Potsdam, Germany
             \and
           Astrophysics Research Institute, Liverpool John Moores University, 
           Liverpool Science Park ic2, 146 Brownlow Hill, Liverpool L3 5RF, UK \\
}
\date{}

\abstract{We present spectral classifications from optical
  spectroscopy of 263 massive stars in the north-eastern region of the
  Large Magellanic Cloud. The observed two-degree field includes the
  massive 30~Doradus star-forming region, the environs of SN1987A, and
  a number of star-forming complexes to the south of 30~Dor.  These
  are the first classifications for the majority (203) of the stars
  and include eleven double-lined spectroscopic binaries. The sample
  also includes the first examples of early OC-type spectra
  (AA$\Omega$~30\,Dor\,248 and 280), distinguished by the weakness of
  their nitrogen spectra and by C~{\scriptsize IV} \lam4658 emission.
  We propose that these stars have relatively unprocessed CNO
  abundances compared to morphologically normal O-type stars,
  indicative of an earlier evolutionary phase. From analysis of
    observations obtained on two consecutive nights, we present
  radial-velocity estimates for 233 stars, finding one apparent
    single-lined binary and nine ($>$3$\sigma$) outliers compared to
  the systemic velocity; the latter objects could be runaway stars or
  large-amplitude binary systems and further spectroscopy is
    required to investigate their nature.}

\maketitle

\section{Introduction}
\label{intro}

Our knowledge of the massive-star populations of the Magellanic Clouds
has increased significantly over the past decade, largely via
observations with multi-object spectrographs
\citep[e.g.][]{mo03,eh04,flames,vfts,f09,riots}. Such surveys have
been used to address questions pertaining to stellar evolution
\citep[e.g.][]{mo03,msngr_flames}, wide-area studies of stellar
kinematics \citep[e.g.][]{eh08}, the formation of massive stars in
relative isolation \citep[e.g.][]{l10,b12,o13}, the structure of
stellar clusters \citep[e.g.][]{vhb12a,vhb12b}, the
properties of the interstellar medium in the Clouds
\citep[e.g.][]{w06,vbt13}, and, via multi-epoch observations, the
binary properties of massive stars \citep[e.g.][]{b09,s13,d15}.

The compendia of known spectral types of massive stars in the Clouds
by \citet{azb09,azb10} emphasised the disparity in our knowledge of
the spectral content of the Clouds, with classifications for 5324
stars in the Small Magellanic Cloud (SMC), but only 1750 in the Large
Magellanic Cloud (LMC). Filtering the LMC catalogue from the
Magellanic Clouds Photometric Survey \citep[MCPS, ][]{z04} by a faint
magnitude limit of $V$\,$=$\,15.3\,mag and a colour cut of
$(V-I)$\,$<$\,0.0\,mag to identify likely O- and luminous B-type
objects gives a total of $\sim$10\,000 stars\footnote{In this
  illustrative calculation we adopt an absolute magnitude for a B0
  dwarf of $-$3.6 \citep{wal72}, a modest extinction of
  $A_V$\,$=$\,0.4\,mag \citep[from a typical reddening toward the LMC
  of $E(B-V)$\,$\sim$\,0.13\,mag from][and assuming $R$, the ratio of
  total-to-selective extinction of 3.1]{pm95}, and a distance modulus
  to the LMC of 18.5\,mag \citep{pgg13}.}. The true number of
early-type stars in the LMC will be influenced by the effects of
crowding in the MCPS photometry and the number of interlopers included
via the above filters, but there clearly remains much to learn about
the massive-star content of the LMC, even in light of recent
classifications for 780 O- and B-type stars in 30~Doradus
\citep{w14,bst}.

In this article we present spectroscopy from observations in
early 2006 (P.I. van Loon) to test the capabilities of the then new
AAOmega spectrograph \citep{s04,s06} on the 3.9\,m Anglo-Australian
Telescope (AAT).  Two fields were observed in the LMC, one in the NE
near the massive 30~Doradus star-forming region and a second in the
NW, centred on the N11 region.  Spectra for thirteen of the targets
have been published to date: seven peculiar `nfp' stars \citep{nfp}, a
massive runaway O2-type star \citep{vfts016}, and five eclipsing
binary systems \citep{m14}. Partly motivated by results from the
VLT-FLAMES Tarantula Survey \citep[VFTS,][]{vfts} regarding runaway
stars \citep{vfts016,bst}, we have revisited the AAOmega data in the
NE field to investigate the spectral content and dynamics of massive
stars in this part of the LMC.

In this paper we present spectral classifications for 263 stars in the
north-eastern region of the LMC, only 60 (23\%) of which have previous
classifications (and often from observations of much lower quality
and/or resolution). We describe the characteristics and reductions of
the observations, including the optical photometry of our sources, in
Section~\ref{data}. The spectral classifications of our targets are
discussed in Section~\ref{class}, and estimates of stellar radial
velocities (RVs) are presented in Section~\ref{rvs}, followed by a
short summary in Section~\ref{summary}.  The observations in the field
centred on the N11 region will be presented in a future article.

\section{Observations}\label{data}

AAOmega is a fibre-fed, twin-arm spectrograph, with light separated
into blue and red arms using a dichroic beam splitter \citep{s04,s06}.
Up to 392 objects across a field on the sky that is 2$^\circ$ in diameter 
can be observed simultaneously, using fibres on the prime-focus focal plate
configured with the robotic positioner of the Two-degree Field facility
\citep[2dF,][]{l02}; each fibre has an on-sky aperture of 2$''$.
Targets are selected in the 2$^\circ$ field using the {\sc configure}
software \citep[see][]{l02}.

\begin{figure*}
\begin{center}
\includegraphics[scale=0.75]{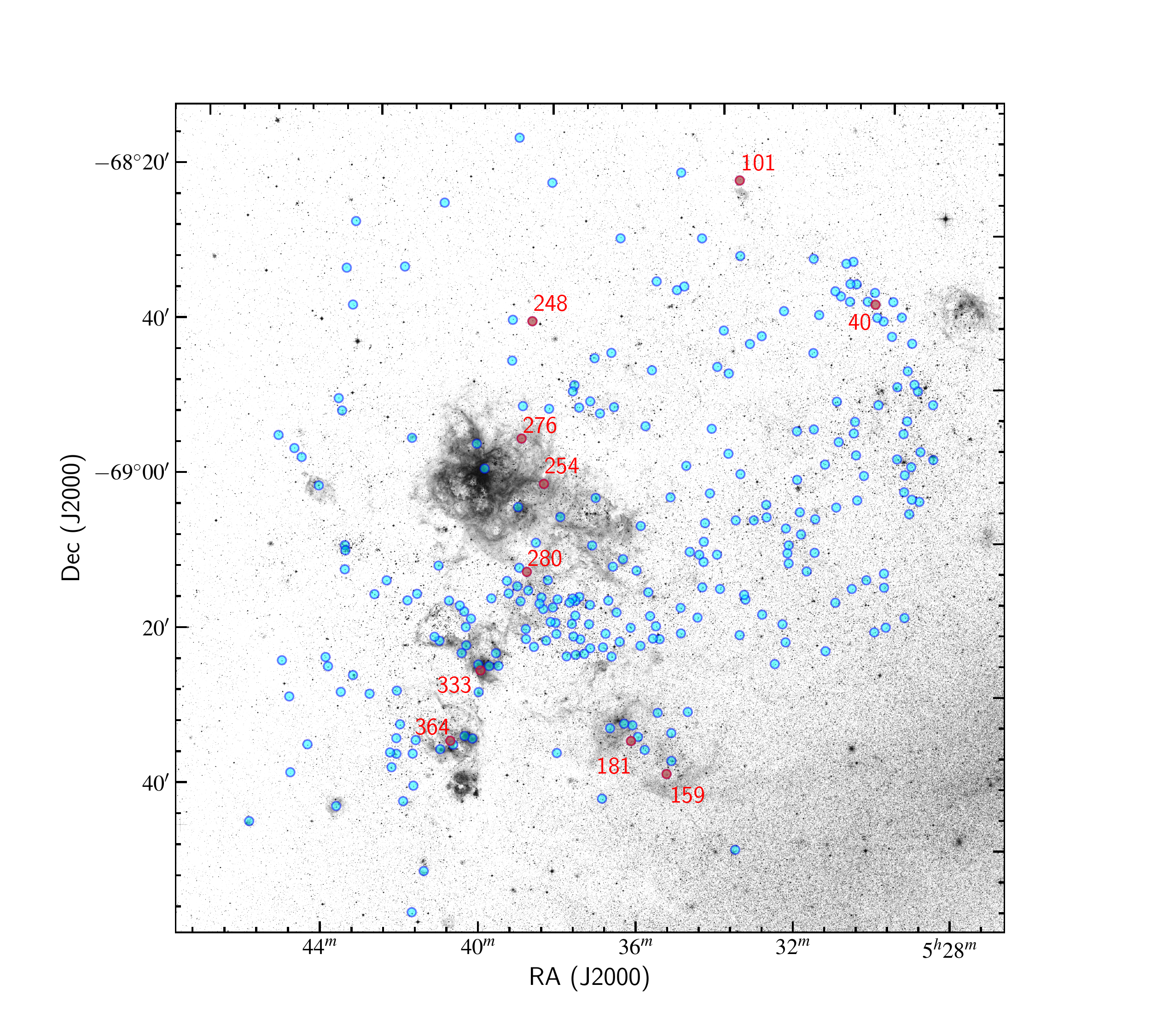}
\caption[]{Location of our AAOmega targets overlaid on a blue-optical
  image from the Digitized Sky Survey (DSS).  The ten early-type stars
  (with classifications of O4 or earlier) discussed in
  Section~\ref{class} are highlighted in red.}\label{targets}
\end{center}
\end{figure*}

Luminous early-type stars were selected as potential targets from the MCPS
\citep{z04} using a magnitude cut of $V$\,$<$\,14\,mag (to ensure
  a signal-to-noise ratio of $>$50) and a colour cut of
$(V-I)$\,$<$\,0.0\,mag to identify early-type stars. We used {\sc
  configure} to select targets from our input list using a field
centre of $\alpha$\,$=$\,05$^{\rm h}$\,36$^{\rm m}$\,07$^{\rm s}$,
$\delta$\,$=$\,$-$69$^\circ$\,15$'$\,58$''$ (J2000). For context,
these coordinates are 3$'$\,NE of the Honeycomb nebula \citep{w92},
3\farcm5\,E of SN1987A, and 17$'$\,SW of R136 (the massive cluster at
the centre of 30~Dor).  The large field (2$^\circ$ diameter) enabled
us to observe a reasonable number of targets in the regions
immediately south of 30~Dor (including NGC\,2060), in the N154, N158,
and N160 complexes further to the south \citep{h56}, and field stars
across the region.  To illustrate the distribution of our targets,
their locations are overlaid on the Digitized Sky Survey
(blue-optical) image in Figure~\ref{targets}.

The NE LMC field was observed on 2006 February 22--23.  The data
presented here were obtained with the blue arm, using the 1700B
grating on the first night, and the 1500V grating on the second.
Simultaneous observations were obtained with the red arm using the
1700D grating and centred at 8620\,\AA; this region contains fewer
lines of interest for massive stars than the blue spectra so
these data were not considered further.

The data were reduced using the {\footnotesize 2}{\sc dfdr} software
\citep{l02}. In brief, {\footnotesize 2}{\sc dfdr} was used for bias
subtraction, fibre location, extractions, division by a normalised
flat-field, and wavelength calibration of each target.  Subsequent
processing included correction of the spectra to the heliocentric
frame, sky subtraction, rejection of significant cosmic rays, and
preliminary normalisation (using pre-defined continuum regions).  The
delivered spectral coverage and resolution from the observations with
the blue arm of the spectrograph is summarised in Table~\ref{obsinfo}.
The signal-to-noise ratio of the final spectra obtained with the 1700B
grating is 50-60 per rebinned pixel for the faintest targets, and in
excess of 100 for the brightest supergiants. The signal-to-noise ratio
of the (longer and slightly lower resolution) 1500V observations is
typically 20-30 greater than for the 1700B data.

Astrometry and optical photometry for each target (from the MCPS) is
listed in Tables~\ref{30dor} and \ref{30dor_phot}, respectively (both
available online) -- the identifiers in the first column are simply
the running numbers (in ascending RA) from our list of potential
targets (hence those observed do not run in a continuous sequence);
for consistency with the identifications used by \citet{nfp} we also
adopt them here.  We note that the majority of the photometry from
\citeauthor{z04} for stars with $V$\,$<$\,13.5\,mag was taken from the
survey by \citet{pm02}, which employed a photometric aperture of
16\farcs2; thus in many instances crowding and/or nebular
contamination may well influence the values in Table~\ref{30dor_phot}
\citep[e.g. see discussion by][]{vfts}. Cross-matches of our targets
with identifications/aliases from past spectroscopy are included in
the final column of Table~\ref{30dor}.

\begin{table}
\caption[]{AAOmega spectrograph settings used.\label{obsinfo}}
\begin{center}
\begin{tabular}{ccccccc}
\hline
Grating & \lam$_{\rm c}$ & \lam-coverage & \multicolumn{2}{c}{$\Delta$\lam (FWHM)} & Exp.\\
& [\AA] & [\AA] & [\AA] & [pix.] & [s]\\
\hline
1700B & 4100 & 3765--4400 & 1.00 & 3.05 & 2$\times$600 \\
1700B & 4700 & 4375--4985 & 1.00 & 3.05 & 2$\times$600 \\
1500V & 4375 & 3975--4755 & 1.25 & 3.25 & 2$\times$900 \\
\hline
\end{tabular}
\end{center}
\end{table}

\section{Spectral classification}\label{class}

The AAOmega spectra were classified by visual inspection in comparison
with standards following the usual precepts for early-type stars
\citep{wf90,gosss,gosss2}, taking into account the reduced metallicity
of the LMC \citep[e.g.][]{f88,wal95,w14,bst}, the effects of
rotational broadening and the spectral resolution of our data. In
brief, the primary diagnostic lines in the O-type spectra are the
ionisation ratios of the helium lines, while also taking into account
absorption from Si~\3 at the latest types.  The additional qualifiers
employed in the classifications are summarised in Table~3 of
\citet{gosss}, and the spectral-type and luminosity-class criteria
used for the later types are those summarised in Tables~4, 5, and 6 of
the same study. The B-type classifications employed the same
spectral-type criteria as those in Tables~1 and 2 of \citet{bst}, with
luminosity classes assigned from the width of the Balmer lines, while
also taking into account the intensity of the silicon absorption lines
at the earlier types; example sequences for O- and B-type spectra were
given by \citet{gosss} and \citet{bst}, respectively.

The classifications of the AAOmega spectra are listed in
Table~\ref{30dor}, representing the first classifications for 203 of
our targets. Previous classifications for the remaining 60 stars are
summarised in Table~\ref{cftypes} (available online and complete to
the best of our knowledge); in many cases the AAOmega spectra are
superior to past spectroscopy, e.g. the resolution of the spectra
obtained by \citet{tn98} was only 8\,\AA. Initial inspection of the
spectra revealed eleven double-lined binaries (SB2s), and a number of
candidate single-lined binaries (SB1s), discussed further in
Section~\ref{binaries}.

Our classifications of the B-type supergiants include suffixes to
indicate nitrogen lines which are strong (Nstr) or weak (Nwk)
compared to morphologically-normal stars of the same adopted spectral
type. These qualifiers were diagnosed from visual inspection of the
CNO absorption features throughout the spectra, but principally
informed by the intensity of the N~\2 \lam3995 and the CNO features in
the \lam\lam4640--4650 region \citep{w76,f91}. An example of the
contrast between Nstr and Nwk spectra of B-type supergiants is given
in Figure~1 of \citet{bst}.

\begin{figure*}
\begin{center}
\includegraphics[angle=0, scale=0.98]{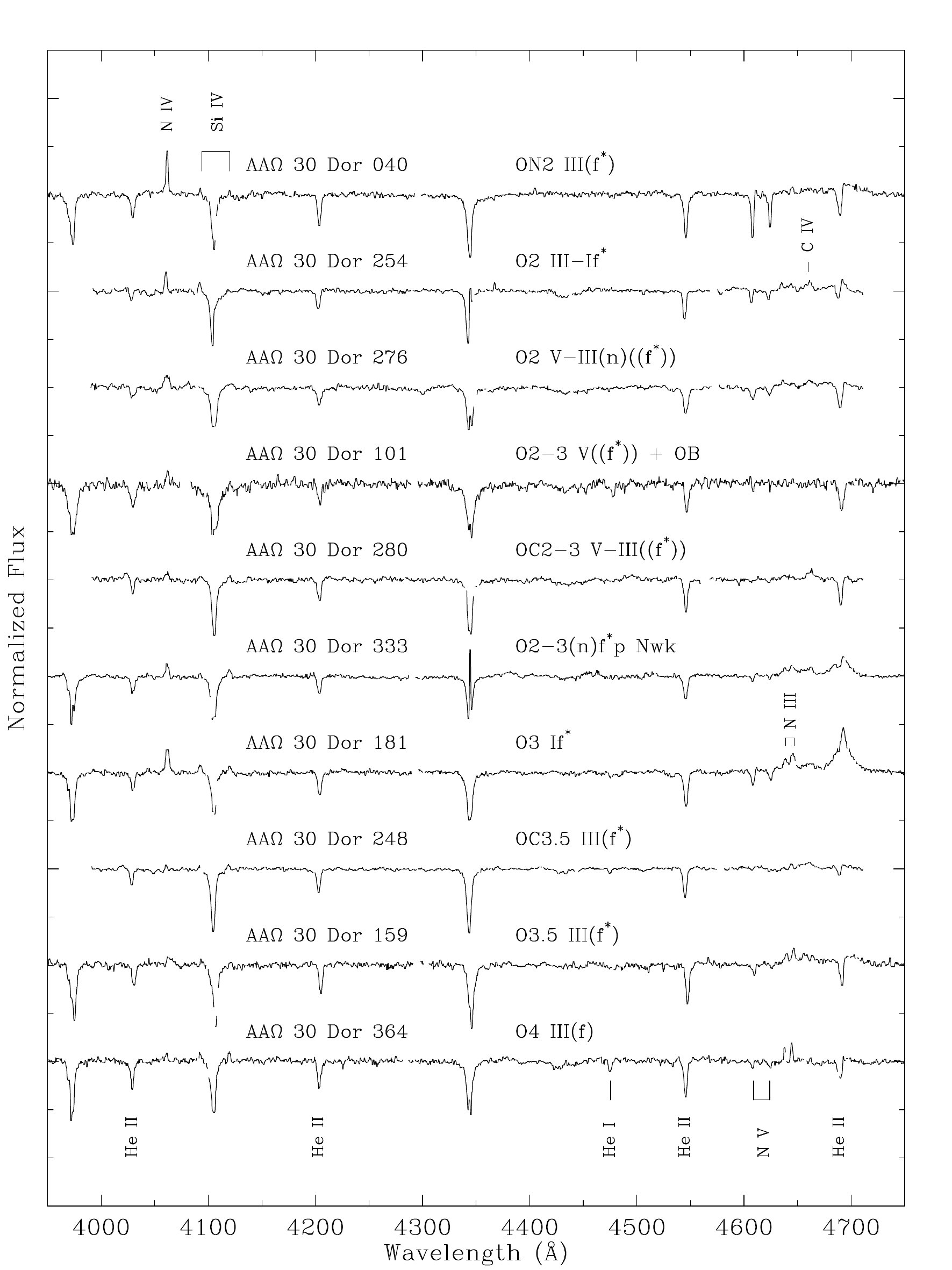}
\caption[]{AAOmega spectra of the ten early O-type stars in the sample
  (with each spectrum smoothed by a 5-pixel median filter for clarity
  and offset by 0.5 continuum units).  Absorption lines identified in
  the spectrum of AA$\Omega$~30\,Dor\,364 are: He~{\scriptsize II}
  \lam\lam4026, 4200, 4542, 4686; He~{\scriptsize I} \lam4471;
  N~{\scriptsize V} \lam\lam4604, 4620.  The emission lines identified
  in the spectra are, in order of increasing wavelength:
  N~{\scriptsize IV} \lam4058; Si~{\scriptsize IV} \lam\lam4089, 4116;
  N~{\scriptsize III} \lam\lam4634-40-42; C~{\scriptsize IV} \lam4658.
  Broad absorption from the \lam4430 diffuse interstellar band can be
  seen in some spectra (e.g.
  AA$\Omega$~30\,Dor\,254).}\label{aao_early}
\end{center}
\end{figure*}

The sample includes ten stars with classifications of O4 or earlier
and their spectra are shown in Figure~\ref{aao_early}. Four of these
were previously unknown, namely: AA$\Omega$~30\,Dor~101, 181, 248, and
280. Given the absence of He~\1 \lam4471 in the spectrum of
AA$\Omega$~30\,Dor\,159, we adopt a classification of
O3.5~III(f$^\ast$) over the O4~III(f) from \citet{w02}. The sample
also includes examples of the nfp class of peculiar O-type
spectra, which are defined by composite emission and absorption in the
He~\2 \lam4686 line \citep{wal73}.  The six nfp stars in the sample,
AA$\Omega$~30\,Dor\,142, 187, 320, 333, 368 and 380, were included in the
discussion of the phenomenon by \citet{nfp}, and we adopt their
classifications here.

The classification of AA$\Omega$\,30\,Dor\,078
\citep[SOI\,399,][]{soi76} also merits brief discussion.  Classified
as A0~Ia by \citeauthor{soi76}, its classification here as B3~Iab is
tantalysing in the context of variations associated with luminous blue
variables. The objective-prism spectroscopy used by \citeauthor{soi76}
was relatively coarse in terms of spectral resolution, but it is
notable that their classifications include spectra as early as B6 (for
comparable magnitudes and from the same photographic plate), so it is
plausible to assume that they would have been able to distinguish
SOI\,399 as a B-type spectrum if it was in the same state as the 2006
observations\footnote{\citet{soi76} cross-matched their star 399 to
  Sk$-$68$^\circ$100, but Brian Skiff's updated catalogues (see
  footnotes to Table~\ref{30dor}) match this \citeauthor{sk69} source
  to SOI\,398 (classified by \citeauthor{soi76} as A1 Ia).}.  Checks
of the ASAS-3 database \citep[spanning 2000--2009, see][]{asas} and
the DASCH archive \citep[spanning 1890--1990,][]{dasch} reveal no
significant photometric variations (given the cadence of the available
data).

\onllongtab{
\footnotesize\onecolumn
\begin{landscape}
\begin{longtable}{lcclccccl}
\caption{Observational parameters of target stars.}\label{30dor}\\
\hline
\hline
Star& $\alpha$\,(J2000) & $\delta$\,(J2000) & Spectral type & $v_1$\,$\pm$\,s.e. [\kms] & $n$ & $v_2$\,$\pm$\,s.e. [\kms] & $n$ & Comments\\
\hline
\endfirsthead
\caption[]{\it{continued}}\\
\hline
Star& $\alpha$\,(J2000) & $\delta$\,(J2000) & Spectral type & $v_1$\,$\pm$\,s.e. [\kms] & $n$ & $v_2$\,$\pm$\,s.e. [\kms] & $n$ & Comments\\
\hline
\endhead
\hline
\multicolumn{9}{r}{\it{continued on next page}}\\
\endfoot
\hline
\endlastfoot
001 & 05 27 33.34 & $-$69 01 22.0 &  O9.7 II-Ib               & 267.0\,$\pm$\,2.7 & 6 & 265.5\,$\pm$\,3.3 &  3 & \\
004 & 05 27 38.31 & $-$69 08 29.9 &  B1 Iab Nwk               & 279.8\,$\pm$\,6.0 & 6 & 277.1\,$\pm$\,4.4 & 6 & \\
008 & 05 27 54.14 & $-$68 59 27.1 &  B0:                      & \ldots  & \ldots &  \ldots &  \ldots & \\
009 & 05 27 56.22 & $-$69 07 21.8 &  B0.7 II                  & 284.6\,$\pm$\,2.7 & 3 & \ldots & \ldots & \\
010 & 05 27 57.98 & $-$68 53 11.9 &  O9.5:~$+$~early B        & \multicolumn{4}{c}{Binary (SB2)} & \\
011 & 05 27 58.44 & $-$68 58 33.1 &  B1.5 Ib Nwk              & 272.8\,$\pm$\,0.7 & 9 & 276.3\,$\pm$\,2.6 & 7 & \\
014 & 05 28 02.57 & $-$69 13 49.5 &  O9.7 Iab                 & 239.2\,$\pm$\,7.3 & 6 & \ldots & \ldots & Sk$-$69$^\circ$144 \\
015 & 05 28 06.98 & $-$68 56 44.4 &  B2 Ib                    & 279.7\,$\pm$\,1.5 & 9 & 278.4\,$\pm$\,2.7 & 7 & Sk$-$68$^\circ$91a \\
016 & 05 28 10.27 & $-$68 49 42.6 &  B9 Ib                    & 282.5\,$\pm$\,1.2 & 4 & 282.3\,$\pm$\,1.4 & 5 & \\
017 & 05 28 11.20 & $-$69 09 14.5 &  B0.7 II-Ib               & 275.1\,$\pm$\,2.5 & 9 & 269.2\,$\pm$\,2.6 & 7 & \\
018 & 05 28 12.60 & $-$69 03 14.7 &  B0.5 Ib Nwk?             & 272.3\,$\pm$\,3.0 & 8 & 278.3\,$\pm$\,5.5 & 7 & \\
019 & 05 28 13.47 & $-$69 13 29.0 &  B1 Ia Nstr               & 285.7\,$\pm$\,1.5 & 9 & 284.9\,$\pm$\,2.1 & 7 & \\
020 & 05 28 18.79 & $-$69 15 19.8 &  O9.5 III(n)              & 273.6\,$\pm$\,6.7 & 6 & \ldots & \ldots & Sk$-$69$^\circ$146 \\
021 & 05 28 19.07 & $-$69 04 52.6 &  B0.5:~$+$~early B        & \multicolumn{4}{c}{Binary (SB2)} \\
024 & 05 28 21.06 & $-$68 47 37.3 &  B1.5 Ib Nwk              & 274.8\,$\pm$\,1.4 & 9 & 287.9\,$\pm$\,2.3 & 7 & \\
025 & 05 28 21.40 & $-$69 10 14.4 &  B1 Ib                    & 281.6\,$\pm$\,2.6 & 9 & 278.0\,$\pm$\,1.8 & 7 & \\
026 & 05 28 23.55 & $-$68 58 43.6 &  B2 Ib                    & 278.8\,$\pm$\,0.8 & 9 & 280.7\,$\pm$\,3.9 & 7 & \\
027 & 05 28 24.06 & $-$69 12 26.2 &  B0.5:~$+$~early B        & \multicolumn{4}{c}{Binary (SB2)} \\
028 & 05 28 26.29 & $-$68 52 07.5 &  B2.5 Ib                  & 281.4\,$\pm$\,1.9 & 9 & 281.5\,$\pm$\,1.3 & 7 & \\
030 & 05 28 31.13 & $-$69 08 06.1 &  B1.5 III-II(n)           &\o266.6\,$\pm$\,10.8 & 4 & 273.4\,$\pm$\,6.7 & 4 & \\
033 & 05 28 36.18 & $-$69 28 47.1 &  B5 Ib                    & 274.5\,$\pm$\,4.0 & 5 & 268.7\,$\pm$\,2.5 & 3 & \\
034 & 05 28 36.93 & $-$68 50 03.1 &  B0.7 Ib(n) Nwk           & 285.2\,$\pm$\,6.1 & 5 & 282.1\,$\pm$\,3.4 & 5 & LH\,64-6 \\
036 & 05 28 45.65 & $-$68 49 31.4 &  B1.5 II-Ib(n)            & \ldots & \ldots &  272.9\,$\pm$\,5.5 & 5 & W61 16-6; LH\,64-7\\
038 & 05 28 46.46 & $-$68 46 15.8 &  B1-1.5 V-III             & \ldots & \ldots & \ldots & \ldots & W61 16-7; LH\,64-33 \\
040 & 05 28 46.97 & $-$68 47 47.9 &  ON2 III(f$^\ast$)       & 273.0\,$\pm$\,4.1 & 3 & 280.2\,$\pm$\,5.3 &  5 & W61 16-8; LH\,64-16 \\
042 & 05 28 52.80 & $-$69 00 53.9 &  B0.7 II-Ib               & 286.4\,$\pm$\,1.8 & 7 & 288.9\,$\pm$\,4.3 & 7 & \\
043 & 05 28 57.68 & $-$68 47 20.4 &  B1 Ib(n) Nwk             & 282.5\,$\pm$\,2.8 & 4 & 288.6\,$\pm$\,2.2 & 4 & W61 16-29 (in LH\,64) \\
045 & 05 29 01.92 & $-$69 22 50.6 &  B1-2e                    & 255.4\,$\pm$\,3.1 & 5 & 265.8\,$\pm$\,3.5 & 4 & \\
047 & 05 29 03.22 & $-$69 24 39.8 &  B9 Ib                    & 287.2\,$\pm$\,4.2 & 4 & 288.2\,$\pm$\,1.6 & 4 & SOI\,624 \\
049 & 05 29 04.89 & $-$69 29 52.4 &  A5 II                    & \ldots & \ldots & 300.0\,$\pm$\,9.6 & 3 & SOI\,625 \\
050 & 05 29 11.90 & $-$68 44 58.8 &  B1 Ib Nwk                & 274.9\,$\pm$\,1.6 & 9 & 275.7\,$\pm$\,2.0 & 7 & W61 16-62; LH\,64-60 \\
053 & 05 29 14.27 & $-$68 42 01.2 &  B0.5 Ib Nwk              & 312.1\,$\pm$\,3.7 & 7 & 256.8\,$\pm$\,3.2 & 6 & \\
054 & 05 29 20.84 & $-$68 44 52.9 &  B0-0.5e                  & \ldots & \ldots & \ldots & \ldots & W61 16-78; LH\,64-63 \\
055 & 05 29 21.20 & $-$69 09 57.3 &  B3 Iab                   & 277.7\,$\pm$\,1.6 & 9 & 274.8\,$\pm$\,1.9 & 5 & \\
056 & 05 29 22.46 & $-$69 30 22.7 &  B1.5 Ib(n)               & 284.8\,$\pm$\,2.1 & 5 & 291.4\,$\pm$\,6.8 & 4 & \\
057 & 05 29 23.21 & $-$68 47 11.0 &  O8 III(n)((f))           & 300.6\,$\pm$\,5.4 & 6 & \ldots & \ldots & LH\,64-4\\
058 & 05 29 24.59 & $-$68 42 12.7 &  B1-1.5 V-III             & 249.9\,$\pm$\,3.4 & 3 & \ldots & \ldots & \\
059 & 05 29 28.58 & $-$69 02 50.5 &  B0.5-0.7 V-III           & \ldots & \ldots & \o293.8\,$\pm$\,11.6 & 3 & \\
060 & 05 29 28.68 & $-$69 23 31.8 &  B1 Ia                    & 257.6\,$\pm$\,1.3 & 9 & 252.4\,$\pm$\,2.4 & 7 & Sk$-$69$^\circ$157 \\
061 & 05 29 30.51 & $-$69 07 12.5 &  B1.5 Ib Nwk              & 282.6\,$\pm$\,2.6 & 9 & 282.7\,$\pm$\,3.6 & 6 & \\
062 & 05 29 31.46 & $-$69 04 20.2 &  B1.5 III-II              & 258.1\,$\pm$\,1.5 & 7 & 251.1\,$\pm$\,4.5 & 7 & \\
063 & 05 29 33.59 & $-$69 13 04.7 &  B1.5 Ib Nwk              & 275.0\,$\pm$\,1.2 & 9 & 276.0\,$\pm$\,0.9 & 7 & \\
064 & 05 29 35.76 & $-$68 46 23.8 &  B1 Iab Nstr              & 285.2\,$\pm$\,1.3 & 9 & 286.1\,$\pm$\,1.8 & 7 & LH\,64-39\\
065 & 05 29 43.21 & $-$68 45 41.2 &  O9.2-9.5 V               & 260.8\,$\pm$\,4.0 & 5 & 265.5\,$\pm$\,5.4 & 3 & \\
068 & 05 29 51.01 & $-$69 24 31.7 &  B3 Ib                    & 272.3\,$\pm$\,3.3 & 9 & 269.9\,$\pm$\,1.5 & 7 & \\
069 & 05 29 52.68 & $-$69 00 02.6 &  B1 II                    & 259.3\,$\pm$\,1.5 & 9 & 273.8\,$\pm$\,2.4 & 7 & \\
071 & 05 29 54.41 & $-$69 05 19.0 &  B3 Ib                    & 270.7\,$\pm$\,2.5 & 7 & 275.1\,$\pm$\,1.0 & 4 & \\
072 & 05 30 05.14 & $-$69 13 47.7 &  B1.5 II-Ib               & 280.8\,$\pm$\,3.8 & 8 & 295.2\,$\pm$\,3.4 & 5 & \\
074 & 05 30 08.99 & $-$68 48 36.6 &  O9.7 Iab                 & 289.3\,$\pm$\,3.9 & 6 & 288.4\,$\pm$\,3.3 & 3 & Sk$-$68$^\circ$105\\
075 & 05 30 10.75 & $-$68 41 15.4 &  B0.5: V-III              &\o296.4\,$\pm$\,16.3 & 3 & \ldots & \ldots & \\
076 & 05 30 16.90 & $-$69 08 06.0 &  B1 Ib                    & 301.3\,$\pm$\,2.7 & 8 & 307.6\,$\pm$\,2.6 & 7  \\
077 & 05 30 16.93 & $-$69 26 09.3 &  B0.2 Ib                  & 282.1\,$\pm$\,3.3 & 8 & 281.2\,$\pm$\,2.6 & 6 & \\
078 & 05 30 21.47 & $-$68 53 30.1 &  B3 Iab                   & 288.0\,$\pm$\,1.0 & 8 & 288.4\,$\pm$\,2.3 & 7 & SOI\,399 \\
079 & 05 30 29.13 & $-$69 03 25.8 &  B0 III                   & 273.4\,$\pm$\,1.9 & 8 & 275.9\,$\pm$\,2.0 & 7 & \\
082 & 05 30 36.99 & $-$69 32 21.4 &  B9 Ib                    & 295.2\,$\pm$\,3.6 & 4 & 297.3\,$\pm$\,5.1 & 4 & \\
083 & 05 30 37.24 & $-$69 15 06.4 &  B1.5 Ia Nstr             & 280.9\,$\pm$\,1.2 & 9 & 275.2\,$\pm$\,1.1 & 7 & Sk$-$69$^\circ$167\\
084 & 05 30 41.93 & $-$69 19 27.9 &  B1.5 Ib Nwk              & 330.8\,$\pm$\,3.1 & 8 & 306.0\,$\pm$\,3.1 & 7 & \\
085 & 05 30 53.95 & $-$69 03 29.7 &  B0 V                     & 310.9\,$\pm$\,3.3 & 5 & 272.9\,$\pm$\,3.3 & 7 & \\
086 & 05 30 55.71 & $-$69 21 48.3 &  B1.5 Ib                  & 272.5\,$\pm$\,1.2 & 9 & 272.7\,$\pm$\,2.6 & 7 & \\
089 & 05 30 58.94 & $-$69 14 02.9 &  B1-1.5 Ib(n)             & 269.0\,$\pm$\,2.0 & 4 & 265.4\,$\pm$\,4.0 & 5 & \\
090 & 05 30 59.11 & $-$68 47 45.3 &  AF (A0kF5)               & \ldots & \ldots & \ldots & \ldots & \\
091 & 05 30 59.28 & $-$69 09 48.6 &  B2.5 Ib(n)               & 277.8\,$\pm$\,3.5 & 7 & 285.5\,$\pm$\,4.3 & 5 & \\
092 & 05 30 59.77 & $-$69 16 56.9 &  B1 Iab                   & 281.5\,$\pm$\,1.6 & 8 & 286.1\,$\pm$\,2.6 & 7 & \\
093 & 05 31 19.23 & $-$69 18 13.8 &  B1 Ib Nwk                & 291.7\,$\pm$\,1.1 & 8 & 290.0\,$\pm$\,2.9 & 7 & \\
094 & 05 31 21.29 & $-$69 16 00.9 &  B1 Iab Nstr?             & 282.9\,$\pm$\,1.9 & 9 & 288.6\,$\pm$\,3.2 & 6 & \\
095 & 05 31 21.38 & $-$69 20 34.5 &  B9 Ib                    & 274.4\,$\pm$\,2.9 & 4 & 277.0\,$\pm$\,2.3 & 4 & \\
096 & 05 31 21.95 & $-$69 19 15.3 &  B1 II                    & 279.9\,$\pm$\,3.6 & 5 & 286.1\,$\pm$\,8.6 & 5 & \\
097 & 05 31 34.00 & $-$68 50 46.7 &  O9 IIIn                  & 209.0\,$\pm$\,7.7 & 4 & 214.7\,$\pm$\,4.7 & 4 & \\
098 & 05 31 35.25 & $-$69 30 47.7 &  B5 Iab                   & 269.9\,$\pm$\,1.5 & 9 & 271.7\,$\pm$\,1.8 & 7 & Sk$-$69$^\circ$176 \\
099 & 05 31 37.74 & $-$69 28 25.5 &  B2 Ib                    & 274.4\,$\pm$\,1.4 & 8 & 270.5\,$\pm$\,2.5 & 7 & \\
100 & 05 31 47.34 & $-$69 12 44.3 &  B0.7 II                  & 275.1\,$\pm$\,1.6 & 7 & 260.1\,$\pm$\,6.9 & 4 & \\
101 & 05 31 47.38 & $-$68 30 19.3 &  O2-3 V((f$^\ast$))\,$+$\,OB & \o309.3\,$\pm$\,12.2 & 3 & 293.5\,$\pm$\,4.1 & 3 & \\
103 & 05 31 48.26 & $-$69 14 21.6 &  B1 Ia                    & 283.4\,$\pm$\,1.6 & 9 & 290.2\,$\pm$\,2.4 & 7 & \\
104 & 05 31 51.87 & $-$68 51 40.1 &  B1.5 Ib                  & 272.6\,$\pm$\,1.2 & 9 & 275.4\,$\pm$\,1.3 & 7 & \\
105 & 05 31 53.96 & $-$69 33 30.1 &  O9.2 II                  & 310.2\,$\pm$\,5.7 & 6 & 307.1\,$\pm$\,3.1 & 4 & Sk$-$69$^\circ$178 \\
106 & 05 31 55.46 & $-$68 40 08.4 &  O6.5 V((f))              & 269.9\,$\pm$\,8.0 & 6 & 288.1\,$\pm$\,6.8 & 3 & D226-5 \citep{o96} \\
109 & 05 32 06.65 & $-$69 26 56.9 &  B1 Ia                    & 273.0\,$\pm$\,1.0 & 8 & 274.9\,$\pm$\,2.2 & 6 & Sk$-$69$^\circ$180 \\
110 & 05 32 07.32 & $-$69 14 36.5 &  B1.5 II                  & 284.7\,$\pm$\,5.8 & 5 & 278.8\,$\pm$\,3.9 & 7 & \\
111 & 05 32 21.29 & $-$69 08 28.0 &  B1 Ib Nwk                & 294.1\,$\pm$\,1.5 & 9 & 298.3\,$\pm$\,3.2 & 7 & \\
112 & 05 32 25.90 & $-$68 55 15.9 &  B1 Iab Nstr              & 264.5\,$\pm$\,1.3 & 9 & 266.7\,$\pm$\,1.7 & 7 & Sk$-$68$^\circ$117 \\
113 & 05 32 27.89 & $-$68 49 38.4 &  O9.7 Iab                 & 320.8\,$\pm$\,3.5 & 5 & 322.6\,$\pm$\,6.0 & 4 & Sk$-$68$^\circ$118 \\
114 & 05 32 29.49 & $-$69 24 48.4 &  B0.2 III:~$+$~early B    & \multicolumn{4}{c}{Binary SB2} \\
115 & 05 32 30.82 & $-$69 24 11.2 &  B1 Ib-Iab Nwk?           & 278.1\,$\pm$\,6.2 & 6 & 282.1\,$\pm$\,4.3 & 3 & \\
116 & 05 32 33.86 & $-$69 14 25.2 &  B1.5 Ib Nwk              & 278.7\,$\pm$\,1.7 & 9 & 280.7\,$\pm$\,4.0 & 7 & \\
117 & 05 32 36.34 & $-$69 05 41.1 &  B0.7 Ib Nwk              & 262.5\,$\pm$\,1.9 & 8 & 269.2\,$\pm$\,3.0 & 7 & \\
120 & 05 32 41.76 & $-$68 54 17.4 &  O9-9.2 III-II(n)         & 305.9\,$\pm$\,5.7 & 5 & 306.2\,$\pm$\,5.1 & 4 & Sk$-$68$^\circ$119 \\
121 & 05 32 42.51 & $-$69 29 21.9 &  B1 Ib                    & 281.0\,$\pm$\,1.6 & 9 & 280.4\,$\pm$\,1.9 & 7 & \\
122 & 05 32 47.88 & $-$68 37 26.1 &  O9.5 III~$+$~B0:         & \multicolumn{4}{c}{Binary (SB2)} & Sk$-$68$^\circ$120 \\
123 & 05 32 57.48 & $-$69 02 15.5 &  B0.7 II                  & 247.7\,$\pm$\,2.9 & 9 & 242.5\,$\pm$\,4.3 & 7 & \\
124 & 05 33 05.67 & $-$69 23 09.1 &  B0.2 III(n)              & \ldots & \ldots & 254.8\,$\pm$\,0.3 & 3 & \\
125 & 05 33 05.70 & $-$69 18 39.9 &  B0.2 III                 & 291.6\,$\pm$\,3.6 & 7 & 296.6\,$\pm$\,1.9 & 6 & \\
126 & 05 33 08.37 & $-$69 10 37.8 &  B8 Iab                   & 270.3\,$\pm$\,2.5 & 5 & 268.5\,$\pm$\,4.4 & 4 & \\
127 & 05 33 09.36 & $-$68 28 39.6 &  O9 III                   & 256.4\,$\pm$\,2.3 & 6 & 255.8\,$\pm$\,0.7 & 4 & \\
128 & 05 33 16.95 & $-$69 57 11.2 &  B5 Ib(n)                 &\o292.6\,$\pm$\,17.3 & 3 & \ldots & \ldots & \\
129 & 05 33 19.13 & $-$68 43 28.8 &  B0 Iab                   & 258.6\,$\pm$\,3.1 & 7 & 259.0\,$\pm$\,0.9 & 6 & \\
130 & 05 33 19.26 & $-$69 14 26.0 &  B0.2:~$+$~early B        & \multicolumn{4}{c}{Binary (SB2)} \\
132 & 05 33 23.37 & $-$69 16 50.3 &  B0.7 Ib                  & 273.9\,$\pm$\,1.2 & 9 & 277.3\,$\pm$\,2.5 & 7 & \\
133 & 05 33 26.37 & $-$69 19 27.6 &  B5 Ib                    & 271.4\,$\pm$\,2.0 & 6 & 272.3\,$\pm$\,3.4 & 5 & \\
134 & 05 33 30.08 & $-$68 43 52.0 &  O9.5 IIIn                & 289.6\,$\pm$\,7.9 & 4 & 296.6\,$\pm$\,7.6 & 3 & \\
135 & 05 33 31.56 & $-$69 22 44.6 &  B1.5 III-II              & 271.6\,$\pm$\,3.0 & 7 & 248.6\,$\pm$\,2.7 & 6 & \\
136 & 05 33 31.71 & $-$69 18 28.6 &  B0 Ia                    & 238.1\,$\pm$\,2.5 & 8 & 244.6\,$\pm$\,1.9 & 5 & Sk$-$69$^\circ$184\\
137 & 05 33 39.26 & $-$69 06 47.1 &  B5 Iab                   & 277.9\,$\pm$\,2.4 & 6 & 274.3\,$\pm$\,2.1 & 6 & \\
138 & 05 33 42.59 & $-$69 26 36.8 &  B1 Ia                    & 270.4\,$\pm$\,2.5 & 8 & 277.7\,$\pm$\,3.8 & 6 & Sk$-$69$^\circ$186\\
139 & 05 33 45.32 & $-$69 17 59.3 &  B0.7 II(n) Nwk           & 279.3\,$\pm$\,5.9 & 4 & 283.5\,$\pm$\,8.4 & 5 & \\
140 & 05 33 58.03 & $-$68 42 29.8 &  O6.5 II(f)               & 280.5\,$\pm$\,3.1 & 6 & \ldots & \ldots & \\
142 & 05 34 06.32 & $-$69 25 09.0 &  O6.5(n)(f)p              & \ldots & \ldots & \ldots & \ldots & BI\,214\\
143 & 05 34 06.35 & $-$69 10 41.1 &  B1 Ib Nstr?              & 287.9\,$\pm$\,1.5 & 8 & 291.8\,$\pm$\,2.6 & 7 & \\
144 & 05 34 09.25 & $-$69 28 28.2 &  O9 II                    & 257.0\,$\pm$\,5.6 & 6 & 264.6\,$\pm$\,6.3 & 4 & \\
145 & 05 34 09.80 & $-$69 38 44.6 &  O9 III(n)                & 285.1\,$\pm$\,4.8 & 6 & 285.7\,$\pm$\,2.5 & 4 & \\
146 & 05 34 16.44 & $-$68 53 58.3 &  O9.5 III                 & 286.0\,$\pm$\,2.8 & 6 & 279.9\,$\pm$\,3.3 & 4 & \\
149 & 05 34 33.17 & $-$69 01 09.5 &  O9.5 III                 & 314.9\,$\pm$\,3.2 & 6 & 304.3\,$\pm$\,2.2 & 5 & \\
150 & 05 34 37.32 & $-$69 41 17.7 &  O6 V((f))z               & 285.5\,$\pm$\,3.1 & 6 & 292.5\,$\pm$\,6.5 & 4 & \\
151 & 05 34 41.01 & $-$69 44 54.1 &  O9.7 V                   & 286.6\,$\pm$\,4.2 & 6 & 294.3\,$\pm$\,6.9 & 3 & LH\,81-1018\\
152 & 05 34 41.55 & $-$69 28 58.0 &  B1.5 Ia                  & 282.6\,$\pm$\,1.6 & 9 & 280.1\,$\pm$\,1.1 & 7 & \\
155 & 05 34 43.86 & $-$68 36 28.7 &  O9.7 II(n)               & 241.7\,$\pm$\,7.7 & 5 & \ldots & \ldots & BI\,220 \\
157 & 05 34 45.24 & $-$69 27 14.9 &  O9.5 III-II              & 269.9\,$\pm$\,3.1 & 6 & 265.7\,$\pm$\,1.7 & 4 & \\
159 & 05 34 50.16 & $-$69 46 32.2 &  O3.5 III(f$^\ast$)      & 370.1\,$\pm$\,3.3 & 3 & 377.8\,$\pm$\,4.1 & 3 & W61 28-23 (in LH\,81) \\
160 & 05 34 51.44 & $-$69 28 48.2 &  B1.5 V-III               &\o288.8\,$\pm$\,10.8 & 3 & 267.7\,$\pm$\,3.3 & 3 & \\
161 & 05 34 51.89 & $-$69 22 44.3 &  B2.5 Ib                  & 273.2\,$\pm$\,1.0 & 9 & 278.3\,$\pm$\,1.3 & 7 & BI\,217 \\
162 & 05 34 52.49 & $-$69 25 50.6 &  B1 Ib                    & 271.7\,$\pm$\,0.8 & 9 & 271.2\,$\pm$\,2.6 & 7 & \\
163 & 05 34 53.74 & $-$69 14 02.6 &  O5 III(f)                & 294.6\,$\pm$\,8.5 & 3 & 307.1\,$\pm$\,8.6 & 3 & Sk$-$69$^\circ$195 \\
164 & 05 34 54.88 & $-$69 38 29.5 &  B0 Ia$^{+}$             & 296.4\,$\pm$\,1.7 & 8 & 301.1\,$\pm$\,2.2 & 6 & Sk$-$69$^\circ$196 \\
168 & 05 35 06.17 & $-$69 19 47.3 &  B2.5 III                 & 271.0\,$\pm$\,3.1 & 5 & 267.0\,$\pm$\,4.2 & 4 & \\
169 & 05 35 10.99 & $-$69 29 35.0 &  O9.2 V                   & 283.3\,$\pm$\,3.3 & 5 & 278.9\,$\pm$\,4.1 & 3 & \\
170 & 05 35 12.50 & $-$68 51 13.4 &  B0.2 III(n)              & 284.0\,$\pm$\,4.2 & 4 & 293.0\,$\pm$\,8.4 & 5 & W61 27-1 (in LH\,85) \\
172 & 05 35 16.18 & $-$68 58 17.9 &  B0.2 III                 & 261.0\,$\pm$\,3.2 & 5 & 277.1\,$\pm$\,2.9 & 6 & LH\,89-7 \\
173 & 05 35 19.37 & $-$69 43 08.6 &  B0.5 Ib Nwk?             & \multicolumn{4}{c}{Binary (SB2?)} \\
174 & 05 35 22.88 & $-$69 27 07.4 &  B0 III                   & 273.6\,$\pm$\,2.6 & 9 & 267.1\,$\pm$\,3.3 & 7 & \\
176 & 05 35 24.46 & $-$69 18 07.6 &  B1 II-Ib                 & 281.3\,$\pm$\,1.1 & 8 & 286.5\,$\pm$\,1.1 & 7 & \\
177 & 05 35 27.55 & $-$69 41 22.2 &  B0.5 III                 & 274.3\,$\pm$\,2.8 & 6 & 279.0\,$\pm$\,2.2 & 7 & \\
178 & 05 35 34.18 & $-$69 39 47.8 &  O6 V((f))                & 273.6\,$\pm$\,1.8 & 3 & 272.0\,$\pm$\,1.6 & 3 & \\
179 & 05 35 37.36 & $-$68 58 56.1 &  B0.2 III(n)              & \ldots & \ldots & \ldots & \ldots & ST92 3-08 (in LH\,89) \\
180 & 05 35 37.40 & $-$68 51 42.6 &  B1 Iab Nwk               & 288.2\,$\pm$\,2.4 & 9 & 292.2\,$\pm$\,4.0 & 6 & \\
181 & 05 35 38.69 & $-$69 41 48.7 &  O3 If$^\ast$            & 287.6\,$\pm$\,5.0 & 3 & 280.8\,$\pm$\,3.6 & 3 & \\
182 & 05 35 40.31 & $-$69 18 58.2 &  B1 Ib Nwk                & 288.6\,$\pm$\,2.5 & 9 & 285.4\,$\pm$\,0.7 & 7 & \\
183 & 05 35 41.18 & $-$69 28 48.1 &  B1.5 Ia Nstr             & 276.1\,$\pm$\,1.2 & 9 & 275.5\,$\pm$\,1.4 & 7 & Sk$-$69$^\circ$208 \\
184 & 05 35 41.58 & $-$69 24 57.6 &  B1 Ib Nwk                & 264.7\,$\pm$\,2.5 & 7 & 269.1\,$\pm$\,2.7 & 6 & \\
185 & 05 35 46.46 & $-$69 39 28.7 &  O9.2 III(n)              & 279.7\,$\pm$\,6.8 & 5 & 285.9\,$\pm$\,7.2 & 3 & \\
186 & 05 35 49.80 & $-$68 57 15.0 &  O7 III((f))              & 299.2\,$\pm$\,4.5 & 5 & 299.4\,$\pm$\,4.8 & 4 & LH\,89-62\\
187 & 05 35 51.92 & $-$69 23 19.0 &  O6n(f)p                  & \ldots & \ldots &  \ldots &  \ldots & \\
189 & 05 35 55.29 & $-$69 30 37.5 &  B1 Ia$^+$                & 250.4\,$\pm$\,1.0 & 9 & 258.3\,$\pm$\,2.1 & 6 & Sk$-$69$^\circ$209 \\
190 & 05 35 55.74 & $-$69 09 51.0 &  B0.5-1 Ia                & 303.7\,$\pm$\,3.6 & 7 & 309.9\,$\pm$\,9.5 & 5 & \\
192 & 05 36 00.80 & $-$69 27 34.9 &  O8.5 IIn                 & \o259.0\,$\pm$\,10.9 & 6 & 279.8\,$\pm$\,6.2 & 4 & \\
194 & 05 36 06.65 & $-$69 29 18.6 &  B0 Ia                    & 254.4\,$\pm$\,2.7 & 9 & 253.6\,$\pm$\,7.6 & 6 & \\
195 & 05 36 06.66 & $-$68 57 54.5 &  B1 Ib Nwk                & 328.3\,$\pm$\,4.1 & 9 & 329.9\,$\pm$\,1.8 & 6 & W61 27-31; LH\,89-68; ST92 3-62 \\
196 & 05 36 07.71 & $-$69 15 57.9 &  O5 V((f))z               & 263.4\,$\pm$\,0.7 & 3 & 258.0\,$\pm$\,3.1 & 3 & \\
197 & 05 36 08.02 & $-$69 39 53.6 &  O9.5 III                 & 279.9\,$\pm$\,6.5 & 4 &\o261.8\,$\pm$\,14.2 & 3 & \\
199 & 05 36 10.16 & $-$68 54 56.4 &  B1.5 Ib Nwk              & 296.2\,$\pm$\,1.8 & 9 & 297.0\,$\pm$\,1.4 & 6 & LH\,89-96 \\
200 & 05 36 12.98 & $-$68 28 24.9 &  B0.7 Ib                  & 283.4\,$\pm$\,1.5 & 9 & 282.9\,$\pm$\,2.7 & 6 & Sk$-$68$^\circ$127\\
201 & 05 36 13.43 & $-$68 55 44.1 &  O9.7-B0 II               & 261.9\,$\pm$\,4.0 & 5 & \ldots & \ldots & W61 27-41 \\
206 & 05 36 19.56 & $-$69 23 38.6 &  B0 III                   & \ldots & \ldots & 253.3\,$\pm$\,8.0 & 3 & \\
207 & 05 36 23.86 & $-$69 26 08.3 &  B0.2 III(n)              & \ldots & \ldots &  \ldots & \ldots & \\
208 & 05 36 25.64 & $-$69 29 15.9 &  O9 IV                    & 245.4\,$\pm$\,4.8 & 6 & 243.7\,$\pm$\,3.0 & 4 & \\
214 & 05 36 30.72 & $-$69 48 54.9 &  O6.5 III(f)              & 243.9\,$\pm$\,3.8 & 6 & 248.9\,$\pm$\,6.9 & 4 & Sk$-$69$^\circ$216a \\
217 & 05 36 34.01 & $-$69 22 28.1 &  O9 II                    & 275.2\,$\pm$\,7.0 & 6 & 272.2\,$\pm$\,8.0 & 4 & \\
218 & 05 36 35.50 & $-$69 29 54.7 &  O9 V                     & 279.3\,$\pm$\,0.8 & 6 & 263.3\,$\pm$\,4.9 & 4 & \\
222 & 05 36 38.92 & $-$69 27 58.8 &  O7.5 Iaf                 & 270.2\,$\pm$\,3.7 & 5 & 262.7\,$\pm$\,2.2 & 3 & \\
226 & 05 36 40.56 & $-$69 22 56.9 &  O7.5 V((f))              & 244.6\,$\pm$\,1.8 & 6 & 244.4\,$\pm$\,0.7 & 4 & \\
228 & 05 36 42.88 & $-$69 24 49.9 &  O9.7-B0 V-III(n)         & 285.4\,$\pm$\,5.6 & 4 & \o292.4\,$\pm$\,11.6 & 3 & \\
229 & 05 36 44.28 & $-$69 22 35.4 &  B0.2 Ia                  & 260.6\,$\pm$\,1.1 & 6 & 257.2\,$\pm$\,1.2 & 4 & \\
232 & 05 36 48.02 & $-$69 29 55.7 &  O9.2 III                 & 274.8\,$\pm$\,3.4 & 6 & 272.4\,$\pm$\,1.9 & 4 & \\
233 & 05 36 48.68 & $-$69 27 30.9 &  B0 Ib                    & 268.0\,$\pm$\,1.4 & 9 & 268.7\,$\pm$\,1.6 & 6 & \\
235 & 05 36 49.19 & $-$69 25 51.8 &  O9.5 III                 & 269.5\,$\pm$\,0.8 & 5 & 262.3\,$\pm$\,1.4 & 4 & \\
236 & 05 36 49.28 & $-$69 23 04.2 &  B0.5 Ib Nwk              & 245.8\,$\pm$\,4.1 & 7 & 239.4\,$\pm$\,7.7 & 5 & \\
237 & 05 36 50.07 & $-$69 11 51.1 &  O7.5 III                 & 277.4\,$\pm$\,1.9 & 5 & 284.2\,$\pm$\,3.9 & 3 & \\
239 & 05 36 50.22 & $-$68 57 40.5 &  A3 Ib                    & 272.0\,$\pm$\,4.6 & 3 & 274.7\,$\pm$\,3.5 & 4 & \\
241 & 05 36 52.66 & $-$68 22 08.8 &  B0 III(n)                &\o243.9\,$\pm$\,15.5 & 3 & \o247.6\,$\pm$\,14.9 & 3 & Sk$-$68$^\circ$132 \\
248 & 05 37 01.34 & $-$68 46 06.2 &  OC3.5 III(f*)            & 230.4\,$\pm$\,9.1 & 3 & 218.5\,$\pm$\,2.0 & 3 & Sk$-$68$^\circ$133 \\
250 & 05 37 01.92 & $-$69 30 00.1 &  B1.5 II                  & 282.8\,$\pm$\,1.8 & 9 & 269.3\,$\pm$\,1.5 & 6 & \\
252 & 05 37 06.53 & $-$69 22 29.6 &  B0.5 II(n)               & \ldots & \ldots & \ldots & \ldots  & \\
254 & 05 37 08.98 & $-$69 07 20.4 &  O2 III-If$^\ast$        & \ldots & \ldots & 190.3\,$\pm$\,0.9 & 3 & VFTS\,016 \\
256 & 05 37 13.21 & $-$69 25 30.6 &  B0.5 Ia Nwk              & 277.7\,$\pm$\,1.6 & 9 & 271.8\,$\pm$\,1.4 & 5 & \\
257 & 05 37 13.55 & $-$69 26 57.7 &  O9.7 II-Ib               & 282.0\,$\pm$\,7.8 & 4 &\o274.9\,$\pm$\,16.4 & 3 & \\
258 & 05 37 14.70 & $-$69 23 28.8 &  O9.5 III(n)              & 281.8\,$\pm$\,9.3 & 3 & \ldots & \ldots & \\
260 & 05 37 17.92 & $-$69 19 52.2 &  B0.7 II                  & 304.5\,$\pm$\,3.4 & 7 & \ldots & \ldots & \\
261 & 05 37 20.22 & $-$69 25 19.6 &  B0.5 II                  & 265.9\,$\pm$\,3.4 & 8 & 280.5\,$\pm$\,5.9 & 5 & \\
264 & 05 37 27.67 & $-$68 56 58.8 &  O8.5 III(n)              & 237.2\,$\pm$\,4.4 & 6 & 233.3\,$\pm$\,3.4 & 4 & BI\,251\\
266 & 05 37 29.07 & $-$68 45 39.2 &  B0.5 II-Ib               & 278.8\,$\pm$\,3.0 & 6 & 277.2\,$\pm$\,4.1 & 4 & \\
267 & 05 37 29.09 & $-$69 23 33.8 &  O9.5: II~$+$~early B     & \multicolumn{4}{c}{Binary (SB2)} & \\
271 & 05 37 29.68 & $-$69 14 51.8 &  O5.5 Iaf                 & \ldots & \ldots & \ldots & \ldots & \\
272 & 05 37 29.72 & $-$69 27 41.1 &  O9.5 Ia                  & 260.2\,$\pm$\,4.2 & 6 & 253.7\,$\pm$\,2.5 & 3 & \\
273 & 05 37 29.93 & $-$69 22 02.0 &  B1.5 Ib Nwk              & 284.3\,$\pm$\,0.6 & 8 & 266.9\,$\pm$\,1.6 & 7 & \\
274 & 05 37 31.35 & $-$69 42 23.4 &  B1 Iab Nstr?             & 241.1\,$\pm$\,1.3 & 9 & 241.0\,$\pm$\,0.8 & 6 & Sk$-$69$^\circ$232 \\
275 & 05 37 33.73 & $-$69 22 48.0 &  O7 V((f))                & 266.1\,$\pm$\,1.8 & 6 &  262.6\,$\pm$\,0.9 & 3 & \\
276 & 05 37 34.45 & $-$69 01 10.0 &  O2 V-III(n)((f$^\ast$)) & \ldots & \ldots & \ldots & \ldots & BI\,253; VFTS\,072 \\
277 & 05 37 36.20 & $-$68 50 56.1 &  B1 Ib-Iab(n) Nwk?        & 293.6\,$\pm$\,3.8 & 4 & 293.8\,$\pm$\,6.2 & 3 & Sk$-$68$^\circ$134 \\
280 & 05 37 47.47 & $-$69 18 30.5 &  O2-3 V-III((f$^\ast$))  & 279.7\,$\pm$\,1.8 & 3 & 276.1\,$\pm$\,2.3 & 3 & \\
281 & 05 37 48.18 & $-$69 20 56.7 &  B1 Ib Nwk                & 300.9\,$\pm$\,1.6 & 9 & 298.6\,$\pm$\,3.2 & 5 & \\
282 & 05 37 48.68 & $-$69 28 19.9 &  O9 II                    & 325.9\,$\pm$\,4.3 & 6 & 314.7\,$\pm$\,4.9 & 4 & \\
283 & 05 37 50.03 & $-$69 09 59.8 &  O7-8 III((f))            & 270.9\,$\pm$\,1.4 & 5 & 282.4\,$\pm$\,6.2 & 3 & ST92 1-72; VFTS\,171 \\
284 & 05 37 57.89 & $-$69 17 52.7 &  O8.5 II((f))             & 265.3\,$\pm$\,3.7 & 6 & 268.8\,$\pm$\,1.5 & 4 & \\
285 & 05 37 57.96 & $-$69 25 54.3 &  O9.7-B0 Ib-Iab           & 316.4\,$\pm$\,1.6 & 4 & 314.7\,$\pm$\,6.1 & 4 & \\
286 & 05 37 59.18 & $-$69 27 12.2 &  O7.5 Ib(f)               & 273.8\,$\pm$\,5.9 & 5 & 269.6\,$\pm$\,5.9 & 3 & Sk$-$69$^\circ$238\\
287 & 05 38 01.27 & $-$69 22 13.9 &  B1 Ia Nstr               & 264.0\,$\pm$\,1.6 & 9 & 269.6\,$\pm$\,2.3 & 6 & Sk$-$69$^\circ$237\\
288 & 05 38 03.80 & $-$69 20 14.7 &  O8.5-9 III               & 268.9\,$\pm$\,2.7 & 6 & 273.6\,$\pm$\,4.2 & 4 & \\
290 & 05 38 17.51 & $-$69 21 02.9 &  B0.7 Ib Nwk              & 250.5\,$\pm$\,8.3 & 6 & 238.7\,$\pm$\,3.6 & 4 & \\
291 & 05 38 18.02 & $-$69 19 24.4 &  O9.7 II                  & 265.1\,$\pm$\,5.0 & 5 & \ldots & \ldots & \\
296 & 05 38 32.76 & $-$69 04 32.8 &  O9 III(n)                & \ldots & \ldots & \ldots & \ldots & P93-304/VFTS\,389\\
308 & 05 38 40.16 & $-$69 01 12.0 &  O8.5-9 II                &\o257.7\,$\pm$\,12.9 & 4 & 269.4\,$\pm$\,3.3 & 4 & P93-9017/VFTS\,481\\
315 & 05 38 43.71 & $-$69 21 27.0 &  B0 Iab                   & \ldots & \ldots & \ldots & \ldots & \\
318 & 05 38 44.26 & $-$69 30 17.6 &  O8 Ib(f)                 & 262.0\,$\pm$\,3.9 & 5 & 277.1\,$\pm$\,3.8 & 4 & ST92 5-85 \\
320 & 05 38 45.98 & $-$69 28 36.9 &  O7n(f)p                  & \ldots & \ldots & \ldots & \ldots & ST92 5-82 \\
324 & 05 38 48.07 & $-$68 29 31.4 &  O9.7-B0 III              & 271.5\,$\pm$\,3.1 & 5 & \ldots & \ldots & Sk$-$68$^\circ$139 \\
330 & 05 38 58.05 & $-$69 30 11.1 &  O9: V-III~$+$~O9.5: V-III& \multicolumn{4}{c}{Binary (SB2)} & W61 3-9; ST92 5-67 \\
333 & 05 39 11.60 & $-$69 30 37.3 &  O2-3(n)f$^\ast$p Nwk    & 267.5\,$\pm$\,8.6 & 3 & \o254.0\,$\pm$\,11.7 & 3 & ST92 5-31 \\
334 & 05 39 13.88 & $-$69 29 49.7 &  O5-6 V((f))z             & 273.3\,$\pm$\,5.4 & 3 & 281.1\,$\pm$\,4.7 & 3 & ST92 5-25 \\
337 & 05 39 17.00 & $-$69 23 46.6 &  O9: V-III~$+$~O9.7: V-III& \multicolumn{4}{c}{Binary (SB2)} &\\
338 & 05 39 17.94 & $-$69 33 24.3 &  O7 V((f))                & 286.5\,$\pm$\,6.8 & 5 & 294.4\,$\pm$\,1.6 & 3 & \\
339 & 05 39 25.84 & $-$69 22 44.9 &  B0.2 Ib                  & 277.4\,$\pm$\,1.7 & 8 & 276.6\,$\pm$\,3.9 & 7 & \\
340 & 05 39 26.04 & $-$69 24 47.1 &  O9.7 II-Ib               & 270.5\,$\pm$\,3.7 & 5 & 266.6\,$\pm$\,5.5 & 4 & \\
341 & 05 39 28.52 & $-$69 27 10.2 &  B1 Ib Nwk                & 260.0\,$\pm$\,3.6 & 7 & 265.0\,$\pm$\,2.5 & 5 & \\
343 & 05 39 31.26 & $-$69 21 54.6 &  O8 II(f)                 & 275.1\,$\pm$\,3.0 & 6 & 278.4\,$\pm$\,7.4 & 4 & \\
344 & 05 39 34.96 & $-$69 39 19.3 &  O6.5-7 III               & \ldots & \ldots & 257.4\,$\pm$\,4.4 & 3 & \\
345 & 05 39 36.58 & $-$69 28 05.3 &  O6.5 II(f)(n)            & \ldots & \ldots & 287.0\,$\pm$\,2.8 & 3 & BI\,260 \\
352 & 05 39 46.20 & $-$69 38 52.7 &  Mid-O V                  & \ldots & \ldots & 258.6\,$\pm$\,6.1 & 3 & F09\,048 \\
353 & 05 39 46.32 & $-$69 21 07.8 &  B0.7 Ib Nwk              & 255.8\,$\pm$\,2.1 & 8 & 256.9\,$\pm$\,2.3 & 6 & \\
355 & 05 39 54.52 & $-$68 37 12.8 &  B0.7 II-Ib               & 269.7\,$\pm$\,1.5 & 7 & 269.1\,$\pm$\,3.7 & 4 & \\
356 & 05 39 55.79 & $-$69 16 25.9 &  B1 Iab  Nstr             & 249.8\,$\pm$\,0.6 & 9 & 245.7\,$\pm$\,3.1 & 6 & Sk$-$69$^\circ$256 \\
361 & 05 40 04.60 & $-$69 39 50.6 &  O5 III(fc)               & \ldots & \ldots & 243.2\,$\pm$\,7.4 & 3 & BI\,265; F09\,082 \\
362 & 05 40 07.15 & $-$69 26 10.7 &  O8.5 III((f))            & 259.5\,$\pm$\,1.6 & 5 & 265.4\,$\pm$\,4.3 & 4 & \\
364 & 05 40 08.21 & $-$69 39 17.1 &  O4 III(f)                & 254.2\,$\pm$\,3.0 & 3 & 262.0\,$\pm$\,2.7 & 3 & F09\,088 \\
367 & 05 40 12.76 & $-$68 59 29.8 &  O5-6 III(f)              & 275.4\,$\pm$\,5.3 & 3 & 270.5\,$\pm$\,2.9 & 3 & BI\,254 \\
368 & 05 40 13.79 & $-$69 25 34.6 &  O7.5n(f)p                & \ldots & \ldots  & \ldots & \ldots & \\
369 & 05 40 24.73 & $-$69 40 13.2 &  O5-6 Vz                  & 240.4\,$\pm$\,4.7 & 3 &\o257.4\,$\pm$\,11.1 & 3 & F09\,111 \\
371 & 05 40 32.03 & $-$69 19 45.7 &  O9.5 II                  & 269.0\,$\pm$\,3.6 & 6 & 269.2\,$\pm$\,1.7 & 3 & \\
374 & 05 40 47.28 & $-$69 20 27.9 &  Early B~$+$~early B      & \multicolumn{4}{c}{Binary (SB2)} &\\
375 & 05 40 56.32 & $-$68 30 35.5 &  B1.5 Ib                  & 288.3\,$\pm$\,0.7 & 9 & 291.0\,$\pm$\,1.5 & 5 & Sk$-$68$^\circ$147 \\
376 & 05 40 59.78 & $-$69 38 40.3 &  B1.5 Ia Nwk              & 249.8\,$\pm$\,1.2 & 9 & 238.9\,$\pm$\,0.5 & 5 & Sk$-$69$^\circ$268; F09\,147 \\
377 & 05 41 06.55 & $-$69 40 22.0 &  B1.5 Ib                  & 254.6\,$\pm$\,1.7 & 9 & 252.6\,$\pm$\,1.3 & 5 & Sk$-$69$^\circ$269; F09\,150 \\
380 & 05 41 11.16 & $-$69 55 44.2 &  O7(n)(f)p                & \ldots & \ldots & \ldots & \ldots & Sk$-$69$^\circ$269a \\
381 & 05 41 11.34 & $-$69 44 31.5 &  O9 III                   & 207.6\,$\pm$\,5.6 & 5 & 206.3\,$\pm$\,9.1 & 4 & F09\,152 \\
383 & 05 41 14.30 & $-$69 17 33.0 &  O9.2 Ib                  & 189.0\,$\pm$\,5.0 & 4 & \ldots & \ldots & \\
384 & 05 41 14.89 & $-$68 41 20.4 &  B3 Ib                    & 283.7\,$\pm$\,2.2 & 9 & 277.4\,$\pm$\,2.2 & 5 & \\
385 & 05 41 17.42 & $-$68 36 28.3 &  A0 II-Ib                 & 258.9\,$\pm$\,7.9 & 4 & 261.2\,$\pm$\,5.6 & 5 & \\
386 & 05 41 18.83 & $-$69 31 58.7 &  B0.5 Ib Nwk              & 254.4\,$\pm$\,5.4 & 5 & 255.5\,$\pm$\,5.4 & 3 & \\
387 & 05 41 20.10 & $-$69 36 22.9 &  B5 Ia                    & 264.0\,$\pm$\,1.2 & 7 & 252.4\,$\pm$\,1.3 & 5  & Sk$-$69$^\circ$271; F09\,158 \\
389 & 05 41 28.16 & $-$69 38 07.1 &  B1 Ib                    & 260.4\,$\pm$\,1.1 & 9 & 263.0\,$\pm$\,1.5 & 6 & F09\,165 \\
390 & 05 41 29.44 & $-$69 46 22.6 &  O9.7 Ia                  & 321.4\,$\pm$\,2.5 & 3 & \ldots & \ldots & F09\,166 \\
391 & 05 41 30.78 & $-$69 40 08.1 &  B0 Ia                    & 267.5\,$\pm$\,0.8 & 8 & 266.2\,$\pm$\,2.2 & 5 & F09\,168 \\
393 & 05 41 34.55 & $-$69 19 08.6 &  B0 Iab                   & 264.1\,$\pm$\,1.6 & 8 & 261.8\,$\pm$\,5.5 & 4 & \\
394 & 05 41 36.78 & $-$70 00 52.4 &  B0.7 Ia                  & 246.6\,$\pm$\,1.2 & 8 & 258.5\,$\pm$\,2.7 & 5 & Sk$-$70$^\circ$111 \\
397 & 05 41 40.12 & $-$69 39 51.6 &  B1 Ib Nwk                & 261.0\,$\pm$\,1.2 & 9 & 264.9\,$\pm$\,2.1 & 6 & F09\,177 \\
398 & 05 41 40.47 & $-$69 41 47.7 &  B0.7 III-II              & 266.3\,$\pm$\,1.9 & 8 & 255.8\,$\pm$\,3.5 & 5 & \\
400 & 05 41 48.89 & $-$68 54 52.5 &  B1.5 Ib Nstr             & 280.6\,$\pm$\,1.8 & 8 & 281.1\,$\pm$\,0.9 & 6 & \\
401 & 05 41 51.52 & $-$68 53 13.5 &  B0.5 Ia                  & 222.5\,$\pm$\,2.5 & 9 & 216.1\,$\pm$\,3.2 & 5 & \\
402 & 05 41 59.77 & $-$69 31 56.7 &  B1 Ib(n) Nwk             &\o245.7\,$\pm$\,13.6 & 4 & 262.9\,$\pm$\,7.6 & 3 & \\
407 & 05 42 09.05 & $-$69 12 19.7 &  O9.5 II                  & 252.8\,$\pm$\,4.0 & 6 & 264.1\,$\pm$\,5.6 & 4 & In NGC\,2100 \\
408 & 05 42 09.05 & $-$69 13 00.4 &  B3 Ia                    & 251.1\,$\pm$\,2.1 & 7 & 250.2\,$\pm$\,1.4 & 6 & In NGC\,2100 \\
409 & 05 42 13.46 & $-$69 15 27.1 &  O7 III((f))              & 313.1\,$\pm$\,2.6 & 6 & 313.9\,$\pm$\,3.7 & 4 & \\
412 & 05 42 20.86 & $-$69 29 16.1 &  A7 II                    & 265.6\,$\pm$\,5.6 & 3 & 262.2\,$\pm$\,7.6 & 3 & \\
414 & 05 42 35.99 & $-$69 04 11.5 &  O8.5 Iabf ($+$\,OB?)     & \multicolumn{4}{c}{Binary (SB2?)} & Sk$-$69$^\circ$287 \\
415 & 05 42 41.93 & $-$69 31 14.4 &  B1 Ib(n)                 & 263.8\,$\pm$\,4.3 & 5 & 256.3\,$\pm$\,5.5 & 4 & \\
416 & 05 42 55.49 & $-$69 00 14.0 &  B0.2-0.5 III:(n)         & \ldots & \ldots & 277.1\,$\pm$\,7.1 & 3 & \\
417 & 05 42 56.01 & $-$69 27 42.8 &  O9.2 II                  & 271.5\,$\pm$\,3.5 & 6 & 281.1\,$\pm$\,8.2 & 4 & Sk$-$69$^\circ$291 \\
419 & 05 42 57.94 & $-$69 26 29.1 &  B1-1.5 I                 & 452.9\,$\pm$\,2.7 & 7 & 430.0\,$\pm$\,2.7 & 3 & \\
421 & 05 43 04.37 & $-$68 58 57.8 &  B1 III                   & 253.3\,$\pm$\,2.1 & 8 & 253.6\,$\pm$\,1.5 & 4 & \\
422 & 05 43 10.49 & $-$69 45 54.8 &  O6-7 Iabf                & \ldots & \ldots & \ldots & \ldots & \\
424 & 05 43 24.79 & $-$68 56 59.7 &  B0.7 Ib  Nwk             & 192.2\,$\pm$\,3.8 & 7 & 199.9\,$\pm$\,2.9 & 6 & Sk$-$68$^\circ$157\\
425 & 05 43 41.34 & $-$69 37 27.4 &  O8 III                   & 270.8\,$\pm$\,3.1 & 5 & 266.3\,$\pm$\,2.5 & 4 & \\
426 & 05 43 58.96 & $-$69 30 57.9 &  O6.5 III(f)              & 213.6\,$\pm$\,6.5 & 5 & 220.7\,$\pm$\,6.9 & 3 & Sk$-$69$^\circ$295 \\
427 & 05 44 02.75 & $-$69 26 10.1 &  B0.7 Iab                 & 243.1\,$\pm$\,3.7 & 9 & 243.9\,$\pm$\,2.8 & 6 & \\
428 & 05 44 12.24 & $-$69 40 46.6 &  O7.5 V((f))              & 166.4\,$\pm$\,2.6 & 5 & 163.7\,$\pm$\,4.1 & 4 & \\
430 & 05 45 23.48 & $-$69 46 22.0 &  O6-7 V                   & \ldots & \ldots & \ldots & \ldots & \\
\hline
\end{longtable}                                                                         
\tablefoot{{\em Cross-references in final column:} \citet[][Sk]{sk69},
  \citet[][BI]{bi75}, \citet[][SOI]{soi76},
  \citet[][ST92{[}1--4{]}-\#\#]{st92}, \citet[][P93]{p93},
  \citet[][ST92 5-\#\#]{tn98}, \citet[][F09]{f09} and
  \citet[][VFTS]{vfts}.  For all but the last two of these references
  we have employed Brian Skiff's updated astrometry for
  cross-identifications (see:
  ftp://ftp.lowell.edu/pub/bas/starcats/\,).  We also include matches
  to stars in LH\,64, LH\,81, LH\,85, and LH\,89 \citep{lh70} from
  \citet{pm00}, in which we include their
  identifications for sources from \citet{w61} where relevant. \\
  {\em Sources of adopted classifications:} {\it
    AA$\Omega$\,30\,Dor\,040:} \citet{w04}; {\it 254:}
  \citet{vfts016}; {\it 276:} \citet{w14};
  {\it 142, 187, 320, 333, 368, \& 380:} \citet{nfp}.\\
  {\em Other relevant comments:} {\it AA$\Omega$\,30\,Dor\,015:} 
  Sk$-$68$^\circ$91a is a pair of stars separated by $\sim$4\farcs5; the AAOmega spectrum is of the southern star.
  {\it AA$\Omega$\,30\,Dor\,283:} Possible contribution from VFTS\,172 at 1\farcs85.
  {\it AA$\Omega$\,30\,Dor\,296:} MCPS position is offset $\sim$0\farcs66 SE of P93-304/VFTS\,389; likely contribution to
observed spectrum as well from P93-294/VFTS\,386 ($\sim$1\farcs4 NW).
  {\it AA$\Omega$\,30\,Dor\,374:} A preliminary classification of B1~III was given by \citet[][identified as lm0031122987]{m14}.}
\end{landscape}
}

\onltab{
\footnotesize\onecolumn
\begin{longtable}{lccccccc}
  \caption{\leftline{MCPS photometry of target stars \citep{z04}.}\label{30dor_phot}}\\
  \hline
  Star & $U$ & $B$ & $V$ & $I_c$ & $U - B$ & $B - V$ & $V - I_c$ \\
  \hline
\endfirsthead
\caption[]{\it{continued}}\\
\hline
  Star & $U$ & $B$ & $V$ & $I_c$ & $U - B$ & $B - V$ & $V - I_c$ \\
\hline
\endhead
\hline
\multicolumn{8}{r}{\it{continued on next page}}\\
\endfoot
\hline
\endlastfoot
\hline
001 & 12.293 & 13.354 & 13.476 & 13.864 &$-$1.061 &$-$0.122 &$-$0.388 \\
004 & 12.635 & 13.372 & 13.486 & 13.729 &$-$0.737 &$-$0.114 &$-$0.243 \\
008 & 12.896 & 14.086 & 13.879 & 14.013 &$-$1.190 &\pp0.207 &$-$0.134 \\
009 & 12.967 & 13.713 & 13.902 & 14.054 &$-$0.746 &$-$0.189 &$-$0.152 \\
010 & 12.303 & 13.274 & 13.436 & 13.786 &$-$0.971 &$-$0.162 &$-$0.350 \\
011 & 12.951 & 13.624 & 13.845 & 13.981 &$-$0.673 &$-$0.221 &$-$0.136 \\
014 & 11.673 & 12.754 & 12.926 & 13.216 &$-$1.081 &$-$0.172 &$-$0.290 \\
015 & 11.763 & 12.734 & 12.836 & 13.363 &$-$0.971 &$-$0.102 &$-$0.527 \\
016 & 12.853 & 13.484 & 13.486 & 13.684 &$-$0.631 &$-$0.002 &$-$0.198 \\
017 & 12.951 & 13.703 & 13.943 & 13.981 &$-$0.752 &$-$0.240 &$-$0.038 \\
018 & 13.031 & 13.780 & 13.832 & 13.862 &$-$0.749 &$-$0.052 &$-$0.030 \\
019 & 12.841 & 13.717 & 13.692 & 13.886 &$-$0.876 &\pp0.025 &$-$0.194 \\
020 & 11.513 & 12.554 & 12.726 & 13.113 &$-$1.041 &$-$0.172 &$-$0.387 \\
021 & 12.593 & 13.564 & 13.706 & 13.801 &$-$0.971 &$-$0.142 &$-$0.095 \\
024 & 12.263 & 13.264 & 13.406 & 13.537 &$-$1.001 &$-$0.142 &$-$0.131 \\
025 & 12.323 & 13.304 & 13.426 & 13.508 &$-$0.981 &$-$0.122 &$-$0.082 \\
026 & 12.483 & 13.364 & 13.436 & 13.553 &$-$0.881 &$-$0.072 &$-$0.117 \\
027 & 12.633 & 13.624 & 13.756 & 13.879 &$-$0.991 &$-$0.132 &$-$0.123 \\
028 & 13.132 & 13.850 & 13.926 & 14.049 &$-$0.718 &$-$0.076 &$-$0.123 \\
030 & 13.088 & 14.152 & 13.853 & 14.285 &$-$1.064 &\pp0.299 &$-$0.432 \\
033 & 13.332 & 13.767 & 13.821 & 13.951 &$-$0.435 &$-$0.054 &$-$0.130 \\
034 & 12.243 & 13.254 & 13.356 & 13.431 &$-$1.011 &$-$0.102 &$-$0.075 \\
036 & 12.889 & 13.823 & 13.883 & 14.101 &$-$0.934 &$-$0.060 &$-$0.218 \\
038 & 12.663 & 13.939 & 13.720 & 14.064 &$-$1.276 &\pp0.219 &$-$0.344 \\
040 & 12.283 & 13.444 & 13.666 & 13.854 &$-$1.161 &$-$0.222 &$-$0.188 \\
042 & 13.039 & 13.682 & 13.804 & 13.857 &$-$0.643 &$-$0.122 &$-$0.053 \\
043 & 12.273 & 13.204 & 13.266 & 13.741 &$-$0.931 &$-$0.062 &$-$0.475 \\
045 & 13.361 & 14.071 & 13.780 & 13.938 &$-$0.710 &\pp0.291 &$-$0.158 \\
047 & 12.843 & 13.214 & 13.216 & 13.294 &$-$0.371 &$-$0.002 &$-$0.078 \\
049 & 13.725 & 13.726 & 13.473 & 13.558 &$-$0.001 &\pp0.253 &$-$0.085 \\
050 & 12.658 & 13.603 & 13.644 & 13.873 &$-$0.945 &$-$0.041 &$-$0.229 \\
053 & 12.491 & 13.511 & 13.534 & 13.829 &$-$1.020 &$-$0.023 &$-$0.295 \\
054 & 12.636 & 13.743 & 13.587 & 13.651 &$-$1.107 &\pp0.156 &$-$0.064 \\
055 & 12.995 & 13.633 & 13.652 & 13.685 &$-$0.638 &$-$0.019 &$-$0.033 \\
056 & 12.543 & 13.474 & 13.576 & 13.812 &$-$0.931 &$-$0.102 &$-$0.236 \\
057 & 12.353 & 13.444 & 13.656 & 13.816 &$-$1.091 &$-$0.212 &$-$0.160 \\
058 & 12.975 & 13.737 & 13.785 & 14.546 &$-$0.762 &$-$0.048 &$-$0.761 \\
059 & 13.059 & 13.869 & 13.858 & 14.169 &$-$0.810 &\pp0.011 &$-$0.311 \\
060 & 11.453 & 12.444 & 12.546 & 12.673 &$-$0.991 &$-$0.102 &$-$0.127 \\
061 & 12.503 & 13.434 & 13.546 & 13.603 &$-$0.931 &$-$0.112 &$-$0.057 \\
062 & 13.013 & 13.654 & 13.778 & 13.906 &$-$0.641 &$-$0.124 &$-$0.128 \\
063 & 12.944 & 13.756 & 13.801 & 13.985 &$-$0.812 &$-$0.045 &$-$0.184 \\
064 & 12.303 & 13.264 & 13.376 & 13.500 &$-$0.961 &$-$0.112 &$-$0.124 \\
065 & 12.581 & 13.841 & 13.911 & 14.127 &$-$1.260 &$-$0.070 &$-$0.216 \\
068 & 13.296 & 13.937 & 13.947 & 14.215 &$-$0.641 &$-$0.010 &$-$0.268 \\
069 & 13.026 & 13.752 & 13.928 & 14.082 &$-$0.726 &$-$0.176 &$-$0.154 \\
071 & 13.230 & 13.844 & 13.841 & 13.902 &$-$0.614 &\pp0.003 &$-$0.061 \\
072 & 12.622 & 13.505 & 13.592 & 13.705 &$-$0.883 &$-$0.087 &$-$0.113 \\
074 & 11.903 & 12.974 & 13.116 & 13.259 &$-$1.071 &$-$0.142 &$-$0.143 \\
075 & 12.836 & 13.709 & 13.888 & 13.908 &$-$0.873 &$-$0.179 &$-$0.020 \\
076 & 12.987 & 13.787 & 13.858 & 13.953 &$-$0.800 &$-$0.071 &$-$0.095 \\
077 & 12.797 & 13.708 & 13.878 & 14.143 &$-$0.911 &$-$0.170 &$-$0.265 \\
078 & 12.293 & 13.094 & 13.186 & 13.210 &$-$0.801 &$-$0.092 &$-$0.024 \\
079 & 12.824 & 13.676 & 13.842 & 14.043 &$-$0.852 &$-$0.166 &$-$0.201 \\
082 & 13.710 & 14.014 & 13.969 & 13.985 &$-$0.304 &\pp0.045 &$-$0.016 \\
083 & 11.343 & 12.304 & 12.386 & 12.510 &$-$0.961 &$-$0.082 &$-$0.124 \\
084 & 12.743 & 13.674 & 13.726 & 14.021 &$-$0.931 &$-$0.052 &$-$0.295 \\
085 & 12.623 & 13.634 & 13.816 & 13.937 &$-$1.011 &$-$0.182 &$-$0.121 \\
086 & 12.513 & 13.434 & 13.546 & 13.884 &$-$0.921 &$-$0.112 &$-$0.338 \\
089 & 13.030 & 13.858 & 13.957 & 13.965 &$-$0.828 &$-$0.099 &$-$0.008 \\
090 & 14.383 & 14.274 & 13.896 & 14.359 &\pp0.109 &\pp0.378 &$-$0.463 \\
091 & 12.583 & 13.494 & 13.596 & 13.687 &$-$0.911 &$-$0.102 &$-$0.091 \\
092 & 12.043 & 13.024 & 13.136 & 13.290 &$-$0.981 &$-$0.112 &$-$0.154 \\
093 & 12.675 & 13.701 & 13.700 & 13.883 &$-$1.026 &\pp0.001 &$-$0.183 \\
094 & 12.323 & 13.314 & 13.436 & 13.590 &$-$0.991 &$-$0.122 &$-$0.154 \\
095 & 13.302 & 13.717 & 13.643 & 13.677 &$-$0.415 &\pp0.074 &$-$0.034 \\
096 & 12.924 & 13.752 & 13.894 & 14.073 &$-$0.828 &$-$0.142 &$-$0.179 \\
097 & 12.799 & 13.709 & 13.872 & 14.172 &$-$0.910 &$-$0.163 &$-$0.300 \\
098 & 12.053 & 12.814 & 12.816 & 12.830 &$-$0.761 &$-$0.002 &$-$0.014 \\
099 & 12.673 & 13.544 & 13.646 & 13.722 &$-$0.871 &$-$0.102 &$-$0.076 \\
100 & 12.802 & 13.590 & 13.784 & 13.931 &$-$0.788 &$-$0.194 &$-$0.147 \\
101 & 13.039 & 14.082 & 13.997 & 14.090 &$-$1.043 &\pp0.085 &$-$0.093 \\
103 & 11.883 & 12.854 & 12.956 & 13.002 &$-$0.971 &$-$0.102 &$-$0.046 \\
104 & 12.744 & 13.897 & 13.647 & 13.778 &$-$1.153 &\pp0.250 &$-$0.131 \\
105 & 12.130 & 13.173 & 13.257 & 13.417 &$-$1.043 &$-$0.084 &$-$0.160 \\
106 & 12.473 & 13.444 & 13.566 & 13.650 &$-$0.971 &$-$0.122 &$-$0.084 \\
109 & 11.853 & 12.844 & 12.976 & 13.134 &$-$0.991 &$-$0.132 &$-$0.158 \\
110 & 13.023 & 13.692 & 13.748 & 13.992 &$-$0.669 &$-$0.056 &$-$0.244 \\
111 & 13.024 & 13.603 & 13.805 & 13.981 &$-$0.579 &$-$0.202 &$-$0.176 \\
112 & 12.443 & 13.414 & 13.526 & 13.552 &$-$0.971 &$-$0.112 &$-$0.026 \\
113 & 11.703 & 12.774 & 12.916 & 13.088 &$-$1.071 &$-$0.142 &$-$0.172 \\
114 & 12.968 & 13.802 & 13.885 & 14.177 &$-$0.834 &$-$0.083 &$-$0.292 \\
115 & 12.413 & 13.414 & 13.526 & 13.689 &$-$1.001 &$-$0.112 &$-$0.163 \\
116 & 12.313 & 13.274 & 13.356 & 13.539 &$-$0.961 &$-$0.082 &$-$0.183 \\
117 & 12.884 & 13.708 & 13.769 & 13.835 &$-$0.824 &$-$0.061 &$-$0.066 \\
120 & 12.473 & 13.534 & 13.676 & 13.758 &$-$1.061 &$-$0.142 &$-$0.082 \\
121 & 12.243 & 13.204 & 13.316 & 13.459 &$-$0.961 &$-$0.112 &$-$0.143 \\
122 & 12.123 & 13.204 & 13.416 & 13.435 &$-$1.081 &$-$0.212 &$-$0.019 \\
123 & 13.175 & 13.899 & 13.901 & 13.948 &$-$0.724 &$-$0.002 &$-$0.047 \\
124 & 12.980 & 13.851 & 13.935 & 14.164 &$-$0.871 &$-$0.084 &$-$0.229 \\
125 & 12.956 & 13.845 & 13.957 & 14.242 &$-$0.889 &$-$0.112 &$-$0.285 \\
126 & 13.237 & 13.676 & 13.580 & 14.436 &$-$0.439 &\pp0.096 &$-$0.856 \\
127 & 12.760 & 13.795 & 13.892 & 14.117 &$-$1.035 &$-$0.097 &$-$0.225 \\
128 & 13.652 & 14.095 & 13.752 & 13.907 &$-$0.443 &\pp0.343 &$-$0.155 \\
129 & 12.113 & 13.134 & 13.246 & 13.355 &$-$1.021 &$-$0.112 &$-$0.109 \\
130 & 12.163 & 13.144 & 13.246 & 13.522 &$-$0.981 &$-$0.102 &$-$0.276 \\
132 & 12.513 & 13.494 & 13.586 & 13.647 &$-$0.981 &$-$0.092 &$-$0.061 \\
133 & 13.109 & 13.689 & 13.623 & 13.738 &$-$0.580 &\pp0.066 &$-$0.115 \\
134 & 12.774 & 13.808 & 13.947 & 14.051 &$-$1.034 &$-$0.139 &$-$0.104 \\
135 & 12.843 & 13.582 & 13.561 & 13.868 &$-$0.739 &\pp0.021 &$-$0.307 \\
136 & 11.923 & 12.924 & 13.046 & 13.161 &$-$1.001 &$-$0.122 &$-$0.115 \\
137 & 12.543 & 13.254 & 13.156 & 13.183 &$-$0.711 &\pp0.098 &$-$0.027 \\
138 & 11.943 & 12.934 & 13.026 & 13.132 &$-$0.991 &$-$0.092 &$-$0.106 \\
139 & 12.799 & 13.642 & 13.619 & 13.906 &$-$0.843 &\pp0.023 &$-$0.287 \\
140 & 12.313 & 13.384 & 13.536 & 13.638 &$-$1.071 &$-$0.152 &$-$0.102 \\
142 & 12.003 & 13.084 & 13.236 & 13.411 &$-$1.081 &$-$0.152 &$-$0.175 \\
143 & 12.563 & 13.474 & 13.566 & 13.606 &$-$0.911 &$-$0.092 &$-$0.040 \\
144 & 12.283 & 13.354 & 13.506 & 13.647 &$-$1.071 &$-$0.152 &$-$0.141 \\
145 & 12.404 & 13.194 & 13.327 & 13.640 &$-$0.790 &$-$0.133 &$-$0.313 \\
146 & 12.860 & 14.104 & 13.930 & 14.089 &$-$1.244 &\pp0.174 &$-$0.159 \\
149 & 12.940 & 13.791 & 13.872 & 13.995 &$-$0.851 &$-$0.081 &$-$0.123 \\
150 & 13.009 & 13.864 & 13.986 & 14.342 &$-$0.855 &$-$0.122 &$-$0.356 \\
151 & 12.954 & 13.603 & 13.743 & 13.927 &$-$0.649 &$-$0.140 &$-$0.184 \\
152 & 11.275 & 11.910 & 11.417 & 12.432 &$-$0.635 &\pp0.493 &$-$1.015 \\
155 & 12.722 & 13.710 & 13.853 & 14.098 &$-$0.988 &$-$0.143 &$-$0.245 \\
157 & 12.737 & 13.616 & 13.793 & 14.023 &$-$0.879 &$-$0.177 &$-$0.230 \\
159 & 12.698 & 13.596 & 13.702 & 13.752 &$-$0.898 &$-$0.106 &$-$0.050 \\
160 & 14.277 & 14.011 & 13.695 & 15.370 &\pp0.266 &\pp0.316 &$-$1.675 \\
161 & 12.563 & 13.384 & 13.416 & 13.500 &$-$0.821 &$-$0.032 &$-$0.084 \\
162 & 12.533 & 13.464 & 13.506 & 13.764 &$-$0.931 &$-$0.042 &$-$0.258 \\
163 & 12.193 & 13.234 & 13.366 & 13.412 &$-$1.041 &$-$0.132 &$-$0.046 \\
164 & 11.323 & 12.163 & 12.159 & 12.545 &$-$0.840 &\pp0.004 &$-$0.386 \\
168 & 12.313 & 13.084 & 13.136 & 14.399 &$-$0.771 &$-$0.052 &$-$1.263 \\
169 & 12.734 & 13.638 & 13.807 & 14.026 &$-$0.904 &$-$0.169 &$-$0.219 \\
170 & 12.820 & 13.533 & 13.752 & 13.969 &$-$0.713 &$-$0.219 &$-$0.217 \\
172 & 13.074 & 13.835 & 13.917 & 14.015 &$-$0.761 &$-$0.082 &$-$0.098 \\
173 & 12.749 & 12.658 & 12.518 & 12.835 &\pp0.091 &\pp0.140 &$-$0.317 \\
174 & 12.339 & 13.188 & 13.287 & 13.544 &$-$0.849 &$-$0.099 &$-$0.257 \\
176 & 13.341 & 14.071 & 13.940 & 14.025 &$-$0.730 &\pp0.131 &$-$0.085 \\
177 & 13.119 & 13.866 & 13.922 & 14.162 &$-$0.747 &$-$0.056 &$-$0.240 \\
178 & 12.507 & 13.349 & 13.297 & 13.726 &$-$0.842 &\pp0.052 &$-$0.429 \\
179 & 13.308 & 14.121 & 13.634 & 13.968 &$-$0.813 &\pp0.487 &$-$0.334 \\
180 & 11.963 & 12.924 & 13.056 & 13.175 &$-$0.961 &$-$0.132 &$-$0.119 \\
181 & 12.335 & 13.139 & 13.280 & 13.514 &$-$0.804 &$-$0.141 &$-$0.234 \\
182 & 13.204 & 13.868 & 13.887 & 14.062 &$-$0.664 &$-$0.019 &$-$0.175 \\
183 & 11.823 & 12.784 & 12.856 & 12.893 &$-$0.961 &$-$0.072 &$-$0.037 \\
184 & 12.951 & 13.743 & 13.824 & 14.065 &$-$0.792 &$-$0.081 &$-$0.241 \\
185 & 12.713 & 13.247 & 13.251 & 13.536 &$-$0.534 &$-$0.004 &$-$0.285 \\
186 & 13.263 & 13.892 & 13.948 & 13.989 &$-$0.629 &$-$0.056 &$-$0.041 \\
187 & 12.113 & 13.214 & 13.396 & 13.481 &$-$1.101 &$-$0.182 &$-$0.085 \\
189 & 10.981 & 11.735 & 11.567 & 11.833 &$-$0.754 &\pp0.168 &$-$0.266 \\
190 & 11.163 & 11.904 & 11.606 & 12.880 &$-$0.741 &\pp0.298 &$-$1.274 \\
192 & 12.579 & 13.507 & 13.520 & 13.809 &$-$0.928 &$-$0.013 &$-$0.289 \\
194 & 11.993 & 13.064 & 13.086 & 13.124 &$-$1.071 &$-$0.022 &$-$0.038 \\
195 & 13.141 & 13.846 & 13.827 & 13.910 &$-$0.705 &\pp0.019 &$-$0.083 \\
196 & 12.948 & 13.905 & 13.870 & 14.117 &$-$0.957 &\pp0.035 &$-$0.247 \\
197 & 14.084 & 13.274 & 13.386 & 13.517 &\pp0.810 &$-$0.112 &$-$0.131 \\
199 & 13.254 & 13.897 & 13.990 & 14.019 &$-$0.643 &$-$0.093 &$-$0.029 \\
200 & 12.593 & 13.604 & 13.726 & 13.896 &$-$1.011 &$-$0.122 &$-$0.170 \\
201 & 13.092 & 13.894 & 13.990 & 14.204 &$-$0.802 &$-$0.096 &$-$0.214 \\
206 & 12.840 & 13.697 & 13.775 & 13.904 &$-$0.857 &$-$0.078 &$-$0.129 \\
207 & 13.062 & 13.957 & 13.937 & 14.334 &$-$0.895 &\pp0.020 &$-$0.397 \\
208 & 11.923 & 12.944 & 13.016 & 13.456 &$-$1.021 &$-$0.072 &$-$0.440 \\
214 & 12.573 & 13.316 & 13.613 & 13.772 &$-$0.743 &$-$0.297 &$-$0.159 \\
217 & 12.273 & 13.304 & 13.426 & 13.473 &$-$1.031 &$-$0.122 &$-$0.047 \\
218 & 12.750 & 13.667 & 13.704 & 14.110 &$-$0.917 &$-$0.037 &$-$0.406 \\
222 & 11.573 & 12.614 & 12.726 & 12.928 &$-$1.041 &$-$0.112 &$-$0.202 \\
226 & 12.836 & 13.833 & 13.847 & 14.080 &$-$0.997 &$-$0.014 &$-$0.233 \\
228 & 13.016 & 13.890 & 13.936 & 14.262 &$-$0.874 &$-$0.046 &$-$0.326 \\
229 & 12.073 & 13.074 & 13.156 & 13.264 &$-$1.001 &$-$0.082 &$-$0.108 \\
232 & 12.330 & 13.163 & 13.166 & 13.433 &$-$0.833 &$-$0.003 &$-$0.267 \\
233 & 13.146 & 14.000 & 13.981 & 14.301 &$-$0.854 &\pp0.019 &$-$0.320 \\
235 & 12.133 & 13.124 & 13.306 & 13.785 &$-$0.991 &$-$0.182 &$-$0.479 \\
236 & 12.969 & 13.819 & 13.917 & 14.049 &$-$0.850 &$-$0.098 &$-$0.132 \\
237 & 12.493 & 13.494 & 13.506 & 13.515 &$-$1.001 &$-$0.012 &$-$0.009 \\
239 & 14.003 & 13.874 & 13.546 & 13.599 &\pp0.129 &\pp0.328 &$-$0.053 \\
241 & 12.811 & 13.799 & 13.939 & 14.169 &$-$0.988 &$-$0.140 &$-$0.230 \\
248 & 12.023 & 13.084 & 13.206 & 13.209 &$-$1.061 &$-$0.122 &$-$0.003 \\
250 & 12.843 & 13.653 & 13.599 & 13.991 &$-$0.810 &\pp0.054 &$-$0.392 \\
252 & 12.859 & 13.693 & 13.592 & 13.747 &$-$0.834 &\pp0.101 &$-$0.155 \\
254 & 12.623 & 13.584 & 13.546 & 13.816 &$-$0.961 &\pp0.038 &$-$0.270 \\
256 & 12.766 & 13.555 & 13.656 & 13.704 &$-$0.789 &$-$0.101 &$-$0.048 \\
257 & 12.353 & 13.344 & 13.416 & 13.553 &$-$0.991 &$-$0.072 &$-$0.137 \\
258 & 12.972 & 13.812 & 13.742 & 13.993 &$-$0.840 &\pp0.070 &$-$0.251 \\
260 & 13.018 & 13.741 & 13.674 & 13.870 &$-$0.723 &\pp0.067 &$-$0.196 \\
261 & 13.004 & 13.818 & 13.913 & 14.004 &$-$0.814 &$-$0.095 &$-$0.091 \\
264 & 12.957 & 13.679 & 13.797 & 13.852 &$-$0.722 &$-$0.118 &$-$0.055 \\
266 & 13.020 & 13.925 & 13.917 & 13.981 &$-$0.905 &\pp0.008 &$-$0.064 \\
267 & 12.223 & 13.204 & 13.356 & 13.423 &$-$0.981 &$-$0.152 &$-$0.067 \\
271 & 12.493 & 13.464 & 13.486 & 13.651 &$-$0.971 &$-$0.022 &$-$0.165 \\
272 & 11.713 & 12.744 & 12.826 & 12.929 &$-$1.031 &$-$0.082 &$-$0.103 \\
273 & 12.663 & 13.544 & 13.516 & 13.539 &$-$0.881 &\pp0.028 &$-$0.023 \\
274 & 12.467 & 13.207 & 13.241 & 13.265 &$-$0.740 &$-$0.034 &$-$0.024 \\
275 & 12.868 & 13.880 & 13.984 & 14.166 &$-$1.012 &$-$0.104 &$-$0.182 \\
276 & 12.765 & 13.650 & 13.669 & 13.742 &$-$0.885 &$-$0.019 &$-$0.073 \\
277 & 11.818 & 13.655 & 13.656 & 13.696 &$-$1.837 &$-$0.001 &$-$0.040 \\
280 & 12.990 & 13.908 & 13.789 & 13.803 &$-$0.918 &\pp0.119 &$-$0.014 \\
281 & 12.860 & 13.760 & 13.655 & 13.895 &$-$0.900 &\pp0.105 &$-$0.240 \\
282 & 12.660 & 13.567 & 13.658 & 13.838 &$-$0.907 &$-$0.091 &$-$0.180 \\
283 & 13.096 & 13.699 & 13.840 & 13.900 &$-$0.603 &$-$0.141 &$-$0.060 \\
284 & 12.917 & 13.787 & 13.724 & 13.780 &$-$0.870 &\pp0.063 &$-$0.056 \\
285 & 12.283 & 13.234 & 13.276 & 13.425 &$-$0.951 &$-$0.042 &$-$0.149 \\
286 & 11.212 & 12.069 & 12.267 & 12.370 &$-$0.857 &$-$0.198 &$-$0.103 \\
287 & 11.243 & 12.122 & 12.032 & 12.107 &$-$0.879 &\pp0.090 &$-$0.075 \\
288 & 13.065 & 13.955 & 13.980 & 14.185 &$-$0.890 &$-$0.025 &$-$0.205 \\
290 & 13.146 & 13.927 & 13.684 & 13.742 &$-$0.781 &\pp0.243 &$-$0.058 \\
291 & 13.227 & 14.056 & 13.920 & 14.168 &$-$0.829 &\pp0.136 &$-$0.248 \\
296 & 12.123 & 13.084 & 12.916 & 13.727 &$-$0.961 &\pp0.168 &$-$0.811 \\
308 & 12.901 & 14.010 & 13.905 & 14.010 &$-$1.109 &\pp0.105 &$-$0.105 \\
315 & 12.393 & 13.384 & 13.466 & 13.556 &$-$0.991 &$-$0.082 &$-$0.090 \\
318 & 12.749 & 13.702 & 13.700 & 13.773 &$-$0.953 &\pp0.002 &$-$0.073 \\
320 & 12.824 & 13.794 & 13.746 & 13.945 &$-$0.970 &\pp0.048 &$-$0.199 \\
324 & 12.713 & 13.637 & 13.801 & 14.087 &$-$0.924 &$-$0.164 &$-$0.286 \\
330 & 12.832 & 13.955 & 13.740 & 14.177 &$-$1.123 &\pp0.215 &$-$0.437 \\
333 & 11.415 & 12.374 & 12.273 & 12.275 &$-$0.959 &\pp0.101 &$-$0.002 \\
334 & 12.810 & 13.966 & 13.551 & 14.168 &$-$1.156 &\pp0.415 &$-$0.617 \\
337 & 13.008 & 13.824 & 13.900 & 14.027 &$-$0.816 &$-$0.076 &$-$0.127 \\
338 & 12.712 & 13.736 & 13.519 & 13.855 &$-$1.024 &\pp0.217 &$-$0.336 \\
339 & 12.912 & 13.767 & 13.821 & 13.891 &$-$0.855 &$-$0.054 &$-$0.070 \\
340 & 11.973 & 13.034 & 13.036 & 13.165 &$-$1.061 &$-$0.002 &$-$0.129 \\
341 & 13.074 & 14.352 & 12.599 & 14.878 &$-$1.278 &\pp1.753 &$-$2.279 \\
343 & 14.855 & 13.826 & 13.771 & 13.985 &\pp1.029 &\pp0.055 &$-$0.214 \\
344 & 13.426 & 13.883 & 13.670 & 13.817 &$-$0.457 &\pp0.213 &$-$0.147 \\
345 & 12.635 & 13.569 & 13.522 & 13.664 &$-$0.934 &\pp0.047 &$-$0.142 \\
352 & 13.216 & 14.168 & 13.787 & 14.096 &$-$0.952 &\pp0.381 &$-$0.309 \\
353 & 13.225 & 14.048 & 13.924 & 14.113 &$-$0.823 &\pp0.124 &$-$0.189 \\
355 & 13.000 & 13.895 & 13.917 & 14.161 &$-$0.895 &$-$0.022 &$-$0.244 \\
356 & 11.713 & 12.644 & 12.666 & 12.772 &$-$0.931 &$-$0.022 &$-$0.106 \\
361 & 11.526 & 12.346 & 12.375 & 12.493 &$-$0.820 &$-$0.029 &$-$0.118 \\
362 & 12.883 & 13.842 & 13.956 & 14.088 &$-$0.959 &$-$0.114 &$-$0.132 \\
364 & 12.867 & 13.562 & 13.629 & 13.637 &$-$0.695 &$-$0.067 &$-$0.008 \\
367 & 13.029 & 13.730 & 13.892 & 14.087 &$-$0.701 &$-$0.162 &$-$0.195 \\
368 & 12.113 & 13.174 & 13.296 & 13.356 &$-$1.061 &$-$0.122 &$-$0.060 \\
369 & 12.013 & 12.820 & 12.365 & 12.638 &$-$0.807 &\pp0.455 &$-$0.273 \\
371 & 12.805 & 13.626 & 13.525 & 13.651 &$-$0.821 &\pp0.101 &$-$0.126 \\
374 & 13.480 & 13.934 & 13.761 & 14.151 &$-$0.454 &\pp0.173 &$-$0.390 \\
375 & 11.923 & 12.894 & 12.966 & 13.088 &$-$0.971 &$-$0.072 &$-$0.122 \\
376 & 11.803 & 12.445 & 12.242 & 12.308 &$-$0.642 &\pp0.203 &$-$0.066 \\
377 & 12.839 & 13.546 & 13.461 & 13.572 &$-$0.707 &\pp0.085 &$-$0.111 \\
380 & 12.573 & 13.416 & 13.444 & 13.557 &$-$0.843 &$-$0.028 &$-$0.113 \\
381 & 12.985 & 13.758 & 13.871 & 14.221 &$-$0.773 &$-$0.113 &$-$0.350 \\
383 & 12.523 & 13.534 & 13.646 & 13.797 &$-$1.011 &$-$0.112 &$-$0.151 \\
384 & 12.959 & 13.698 & 13.628 & 13.671 &$-$0.739 &\pp0.070 &$-$0.043 \\
385 & 13.672 & 13.859 & 13.799 & 13.809 &$-$0.187 &\pp0.060 &$-$0.010 \\
386 & 12.881 & 13.777 & 13.658 & 13.896 &$-$0.896 &\pp0.119 &$-$0.238 \\
387 & 11.307 & 11.790 & 11.790 & 11.910 &$-$0.483 &\pp0.000 &$-$0.120 \\
389 & 12.137 & 12.410 & 12.848 & 13.071 &$-$0.273 &$-$0.438 &$-$0.223 \\
390 & 11.515 & 12.170 & 12.285 & 12.303 &$-$0.655 &$-$0.115 &$-$0.018 \\
391 & 11.673 & 12.479 & 12.584 & 12.945 &$-$0.806 &$-$0.105 &$-$0.361 \\
393 & 12.143 & 13.154 & 13.296 & 13.406 &$-$1.011 &$-$0.142 &$-$0.110 \\
394 & 10.812 & 11.872 & 11.516 & 11.965 &$-$1.060 &\pp0.356 &$-$0.449 \\
397 & 12.795 & 13.507 & 13.693 & 13.755 &$-$0.712 &$-$0.186 &$-$0.062 \\
398 & 13.011 & 13.739 & 13.883 & 14.041 &$-$0.728 &$-$0.144 &$-$0.158 \\
400 & 12.663 & 13.514 & 13.536 & 13.627 &$-$0.851 &$-$0.022 &$-$0.091 \\
401 & 11.963 & 12.954 & 13.036 & 13.104 &$-$0.991 &$-$0.082 &$-$0.068 \\
402 & 13.297 & 14.086 & 13.795 & 14.045 &$-$0.789 &\pp0.291 &$-$0.250 \\
407 & 12.253 & 13.114 & 13.156 & 13.545 &$-$0.861 &$-$0.042 &$-$0.389 \\
408 & 11.893 & 12.614 & 12.566 & 12.919 &$-$0.721 &\pp0.048 &$-$0.353 \\
409 & 11.923 & 13.024 & 13.206 & 13.372 &$-$1.101 &$-$0.182 &$-$0.166 \\
412 & 11.453 & 12.354 & 12.496 & 13.112 &$-$0.901 &$-$0.142 &$-$0.616 \\
414 & 11.333 & 12.224 & 12.216 & 14.113 &$-$0.891 &\pp0.008 &$-$1.897 \\
415 & 13.139 & 13.961 & 13.841 & 13.912 &$-$0.822 &\pp0.120 &$-$0.071 \\
416 & 13.225 & 13.795 & 13.954 & 13.988 &$-$0.570 &$-$0.159 &$-$0.034 \\
417 & 11.653 & 12.664 & 12.696 & 13.130 &$-$1.011 &$-$0.032 &$-$0.434 \\
419 & 12.263 & 13.234 & 13.256 & 13.354 &$-$0.971 &$-$0.022 &$-$0.098 \\
421 & 13.079 & 13.672 & 13.703 & 14.008 &$-$0.593 &$-$0.031 &$-$0.305 \\
422 & 12.531 & 13.468 & 13.321 & 13.364 &$-$0.937 &\pp0.147 &$-$0.043 \\
424 & 12.173 & 13.194 & 13.326 & 13.408 &$-$1.021 &$-$0.132 &$-$0.082 \\
425 & 12.857 & 13.733 & 13.846 & 13.983 &$-$0.876 &$-$0.113 &$-$0.137 \\
426 & 12.572 & 13.549 & 13.702 & 13.826 &$-$0.977 &$-$0.153 &$-$0.124 \\
427 & 13.090 & 13.932 & 13.570 & 13.587 &$-$0.842 &\pp0.362 &$-$0.017 \\
428 & 12.996 & 13.819 & 13.967 & 14.215 &$-$0.823 &$-$0.148 &$-$0.248 \\
430 & 12.969 & 13.947 & 13.991 & 14.215 &$-$0.978 &$-$0.044 &$-$0.224 \\
\hline
\end{longtable}     
}                                                                    

\onltab{
\begin{table*}
\footnotesize
\caption{Comparison of AAOmega classifications with those available in the literature.}\label{cftypes}
\begin{tabular}{llll}
\hline
Star & Alias & AAOmega & Published \\
\hline
034 & LH64-6                                  & B0.7 Ib(n) Nwk & B1 III [M00] \\
036 & W61 16-6, LH64-7                        & B1.5 II-Ib(n) & B1: III: [M00] \\
038 & W61 16-7, LH64-33                       & B1-1.5 V-III & B1 III [M00] \\
040 & W61 16-8, LH64-16                       & ON2 III(f$^\ast$) & O3 III:(f$^\ast$) [M00]; O2 III(f$^\ast$) [W02]; ON2 III(f$^\ast$) [W04] \\
043 & W61 16-29                               & B1 Ib(n) Nwk & B0.2: III: [M00] \\
047 & SOI\,624                                & B9 Ib        & A1 Iab [SOI]\\
049 & SOI\,625                                & A5 II        & B9 Ib [SOI] \\
050 & W61 16-62, LH64-60                      & B1 Ib Nwk & B1.5 III [M00] \\
074 & Sk$-$68$^\circ$105                    & O9.7 Iab & B0 Ia [M95] \\
078 & SOI\,399                                & B3 Iab & A0 Ia [SOI] \\
084 & \ldots                                  & B1.5 Ib Nwk & B1.5 Ib [M14; their lm0020n19615] \\
106 & D226-5                                  & O6.5 V((f)) & O6.5 V((f)) [O96b] \\
112 & Sk$-$68$^\circ$117                    & B1 Iab Nstr & B1 III [M95] \\
120 & Sk$-$68$^\circ$119                    & O9-9.2 III-II(n) & O9 V [M95] \\
142 & BI\,214                                 & O6.5(n)(f)p & O6.5 f [C86]; AAOmega type is from W10 \\
151 & LH\,81-1018                             & O9.7 V & B0.5 III [M00] \\
155 & BI\,220                                 & O9.7 II(n) & B0.5 III [M95] \\
159 & W61 28-23                               & O3.5 III(f$^\ast$) & O3 V((f)) [M00]; O4 III(f$+$) [W02]; O3.5 V((f$+$)) [M05] \\
161 & BI\,217                                 & B2.5 Ib & B2 III [M95] \\
170 & W61 27-1                                & B0.2 III(n) & B0.5 III [M00] \\
179 & ST 3-08                                 & B0.2 III(n) & B0 III [ST92] \\ 
180 & W61 27-7, LH\,85-22                     & B1 Iab Nwk & B1.5: I [M00] \\
187 & \ldots                                  & O6n(f)p & AAOmega type is from W10\\
195 & W61 27-31, LH\,89-68, ST92 3-62         & B1 Ib Nwk & B1 I [ST92] \\
200 & Sk$-$68$^\circ$127                    & B0.7 Ib & B0.5 Ia [M95] \\
214 & Sk$-$69$^\circ$216a                   & O6.5 III(f) & O6 V((f)) [M95] \\
241 & Sk$-$68$^\circ$132                    & B0 III(n) & B0 V [M95] \\
254 & VFTS\,016                              & O2 III-If$^\ast$ & O2 III-If$^\ast$ [E10] \\
276 & BI\,253, VFTS\,072                      & O2 V-III(n)((f$^\ast$)) & O3 V [M95]; O2 V((f$^\ast$)) [W02]; adopted type is from W14 \\
277 & Sk$-$68$^\circ$134                    & B1 Ib-Iab(n) Nwk? & B1 Ib [M95] \\
283 & ST92 1-72, VFTS\,171                    & O7-8 III((f)) & O8 III [ST92]; O8 II-III(f) [W14] \\
286 & Sk$-$69$^\circ$238                    & O7.5 Ib(f) & O6.5 V((f)) [M95] \\
287 & Sk$-$69$^\circ$237                    & B1 Ia Nstr & B1 Ia (N str) [F91] \\
296 & P93-304, VFTS\,389                      & O9 III(n) & O8 V [M85]; O8 V [WB97]; O9.5 V [W02b]; O9.5 IV [W14] \\
308 & P93-9017, VFTS\,481                     & O8.5-9 II & O8.5 V [P93]; O8.5 III [W14] \\
318 & ST92 5-85                               & O8 Ib(f) & O7.5 I(f) [TN98] \\
320 & ST92 5-82                               & O7n(f)p& O6.5 III [TN98]; AAOmega type is from W10 \\ 
330 & W61 3-9, [M2002] LMC\,172231            & O9: V-III\,$+$\,O9.5: V-III & O9.5 III [ST92 5-67, TN98]; O9 V\,$+$\,O9.5 V [M12] \\
333 & ST92 5-31                               & O2-3(n)f$^\ast$p Nwk & O3 If [TN98]; AAOmega type is from W10 (plus addition of Nwk)\\
334 & ST92 5-25                               & O5-6 V((f))z & O4 V [TN98] \\
345 & BI\,260                                 & O6.5 II(f)(n) & O7 V((f)) [M95] \\
352 & F09\,048                                & Mid-O V & O4-6 Vz [F09] \\
361 & BI\,265, F09\,082                       & O5 III(fc) & O6 V [M95]; O5n(f$+$)p [F09] \\ 
364 & F09\,088                                & O4 III(f) & O4 III(f) [F09] \\
367 & BI\,254                                 & O5-6 III(f) & O8 V [M95] \\
368 & \ldots                                  & O7.5n(f)p & AAOmega type is from W10\\
369 & F09\,111                                & O5-6 Vz & O6: Vz [F09] \\
374 & \ldots                                & Early B\,$+$\,Early B & B1 III (SB2) [M14; their lm0031122987] \\
375 & Sk$-$68$^\circ$147                    & B1.5 Ib & B2 II [J01] \\
376 & Sk$-$69$^\circ$268, F09\,147          & B1.5 Ia Nwk & BC1.5 Iab [F09] \\
377 & Sk$-$69$^\circ$269, F09\,150          & B1.5 Ib & B1 III: [M95]; B1.5 Iab [F09] \\
380 & Sk$-$69$^\circ$269a                   & O7(n)(f)p & AAOmega type is from W10\\
381 & F09\,152                                & O9 III & O9 III [F09] \\
387 & Sk$-$69$^\circ$271, F09\,158          & B5 Ia & B4 III-I [F09] \\
389 & F09\,165                                & B1 Ib & B1.5-2 III-II [F09] \\
390 & F09\,166                                & O9.7 Ia & O9.7 II-Ib  [F09] \\
391 & F09\,168                                & B0 Ia & O9.7 Iab [F09] \\
394 & Sk$-$70$^\circ$111                    & B0.7 Ia & B0.5 Ia [F88,F91] \\
397 & F09\,177                                & B1 Ib Nwk & B1.5 Ib [F09] \\
417 & Sk$-$69$^\circ$291                    & O9.2 II & B0 III [C86] \\
\hline 
\end{tabular}
\tablefoot{Sources of classifications: SOI \citep{soi76}; M85 \citep{m85}; C86
  \citep{cgm86}; F88 \citep{f88}; F91 \citep{f91}; M95 \citep{pm95};
  O96b \citep{o96b}; WB97 \citep{wb97}; TN98 \citep{tn98}; M00
  \citep{pm00}; J01 \citep{jaxon}; W02 \citep{w02}; W02b
  \citep{w02hst}; W04 \citep{w04}; M05 \citep{mfast2}; F09 \citep{f09}; E10
  \citep{vfts016}; W10 \citep{nfp}; M12 \citep{pm12}; M14 \citep[][which 
  were earlier classifications of the AAOmega spectra by CJE]{m14}; W14 \citep{w14}.}
\end{table*}
}

\subsection{Indications of CNO abundances in early O-type stars}\label{CNO}

From inspection of the spectra in Figure~\ref{aao_early} we noted the
apparent weakness of the nitrogen features in AA$\Omega$~30\,Dor\,248
and 280. These lines are the primary classification criteria at such
early types \citep[see][]{w02}, yet in these two spectra there is only
weak/marginal N~\3 \lam\lam4634-40-42 and N~\4 \lam4058 emission, and
an absence of N~\5 \lam\lam4604-4620 absorption.  The absence/weakness
of He~\1 \lam4471 (and other He~\1 features) argues for the
early-types adopted here, and we suggest these are the nitrogen-poor
counterparts of the morphologically normal and nitrogen-rich O2-type
stars discussed by \citet{w04}.

Nitrogen enrichment/deficiency in the spectra of late O- and early
B-type spectra was first noted by \citet{w76}, primarily with
reference to the absorption strengths of the CNO features in the
\lam\lam4640--4650 region.  At earlier types, from both morphlogical
considerations and quantitative analysis, \citet{w04} argued that some
O2-type spectra (classified as ON2) are nitrogen-rich compared to
morphologically normal O2-type spectra, suggesting a more advanced
evolutionary state or greater chemical enrichment via initially-larger
rotational velocities.

\begin{figure}
\begin{center}
\includegraphics[scale=0.98]{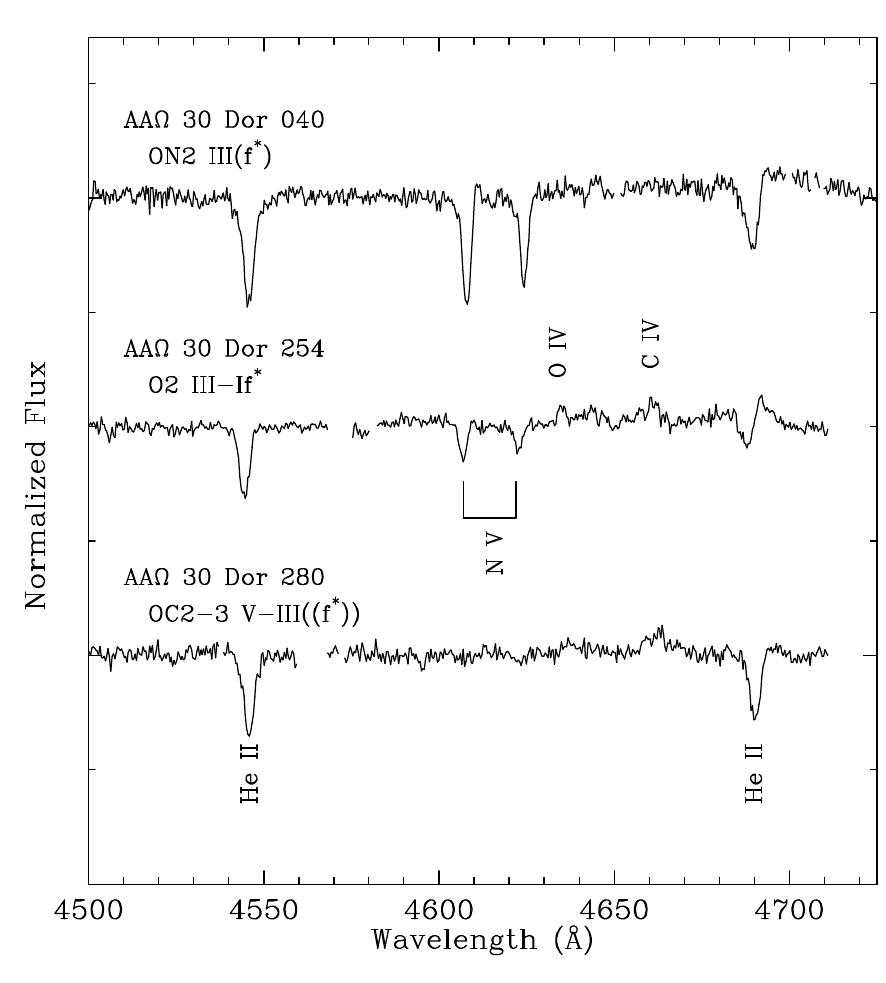}
\caption[]{\lam\lam4500-4725 range of three early-type AAOmega
  spectra, illustrating the large variations in N~{\scriptsize V}
  \lam\lam4604-20 absorption and the presence of C~{\scriptsize IV}
  \lam4658 (and O~{\scriptsize IV} \lam4632) emission.}\label{CN}
\end{center}
\end{figure}

By analogy to the ON/OC sequence at later types
\citep[e.g.][]{w76,gosss}, we therefore classify
AA$\Omega$~30\,Dor\,248 and 280 as OC-type spectra. As well as the
chemical abundances, a broad range of physical factors (e.g. gravity
and mass-loss rate) influence the appearance of the nitrogen features
in the earliest O-type stars \citep{rg12}. On morphological grounds we
employ the OC classification for the two spectra with an absence of
N~\5 \lam\lam4604, 4620 absorption, combined with C~\4 \lam4658
emission. In addition, the weak but still discernable N~\5 absorption
in the spectrum of AA$\Omega$~30\,Dor\,333 argues for an Nwk qualifier
(Walborn, priv.  comm.). As at later types, we suggest that
AA$\Omega$~30\,Dor\,248 and 280 could be examples of an earlier
evolutionary stage (i.e. less chemical processing/enrichment) than the
morphologically-normal objects such as AA$\Omega$~30\,Dor\,254, with
the ON2 star, AA$\Omega$~30\,Dor\,040, completing the sequence. To
highlight the trend in the N~\5 absorption lines in this sequence (and
the presence of C~\4 emission), a subset region of the AAOmega spectra
for three stars is shown in Fig.~\ref{CN}.

Pending further observations (including the H$\alpha$ profiles to
constrain the wind properties), we calculated investigative synthetic
spectra using the PoWR model atmosphere code \citep{hg03,hg04}.
Initially developed for analysis of Wolf--Rayet type spectra, the code
is also well suited for analysis of O-type stars
\citep[e.g.][]{o07,sk183}.  Adopting the physical properties from the
published analysis of VFTS\,016 (i.e. AA$\Omega$~30\,Dor\,254) from
\citet{vfts016}, we calculated seven PoWR models to investigate if the
C~\4 \lam4658 line is genuinely sensitive to abundance, or if it is
significantly affected by other physical parameters.

\begin{figure}
\begin{center}
\includegraphics[scale=0.8]{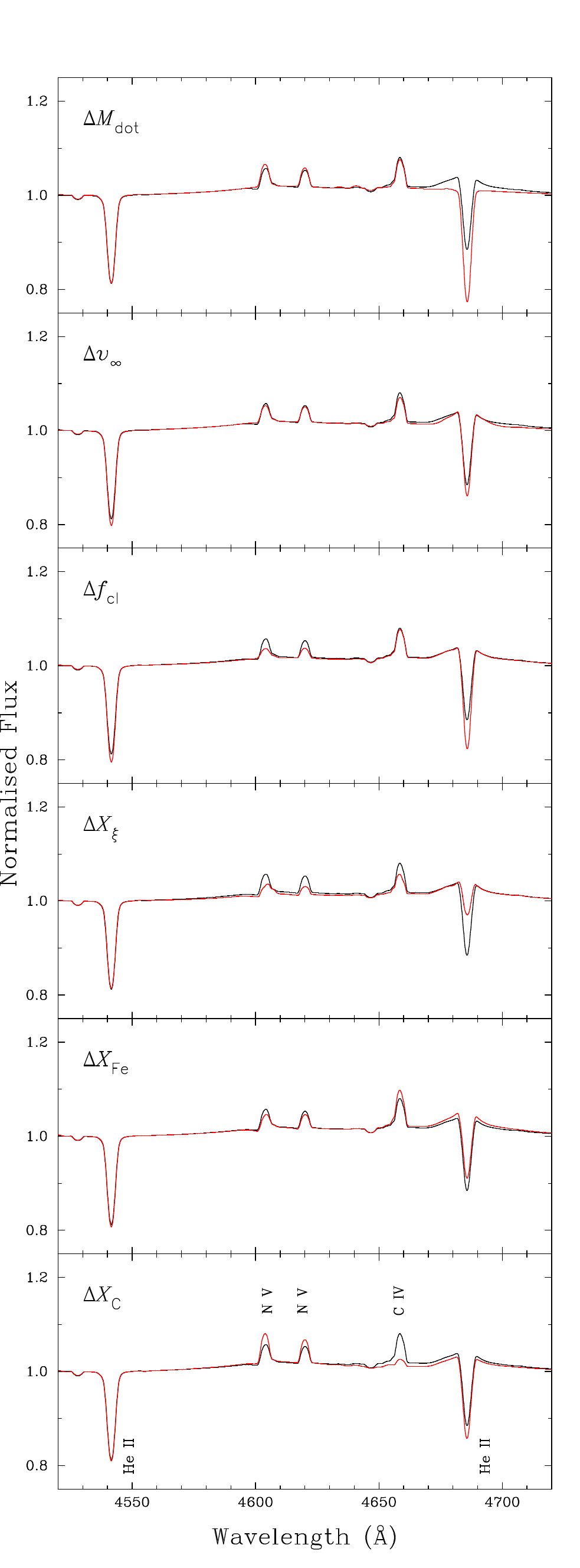}
\caption{Synthetic PoWR spectra (black line) adopting parameters for
  AA$\Omega$~30\,Dor\,254 from \citet{vfts016}.  The spectra plotted
  in red, moving from the top to bottom, are models which vary the
  mass-loss rate ($M_{\rm dot}$), terminal velocity ($v_{\infty}$)
  clumping factor ($f_{\rm cl}$), microturbulence ($\xi$), iron
  abundance ($X_{\rm Fe}$, to investigate possible blanketing
  effects), and carbon abundance ($X_{\rm C}$), as detailed in
  Section~\ref{CNO}. The lines identified in the lower panel are
  He~{\scriptsize II} \lam\lam4542, 4686; N~{\scriptsize V}
  \lam\lam4604-20, and C~{\scriptsize IV} \lam4658.}\label{models}
\end{center}
\end{figure}

The baseline model parameters were: an effective temperature ($T_{\rm
  eff}$) of 50\,kK, luminosity ($L$) of
log\,($L$/$L_\odot$)\,$=$\,6.08, gravity ($g_{\rm grav}$) of
log\,$g_{\rm grav}$\,$=$\,3.75, microturbulence ($\xi$) of 30\,\kms,
and a stellar wind with a terminal velocity ($v_\infty$) of
3450\,\kms, an acceleration law described by a $\beta$ parameter of
1.0, and with a mass-loss rate ($\dot{M}$) of
10$^{-5.5}$\,$M_\odot$\,yr$^{-1}$. Chemical abundances (by mass
fraction) were $X_{\rm H}$\,$=$\,0.7374, $X_{\rm He}$\,$=$\,0.258,
$X_{\rm N}$\,$=$\,0.0008, $X_{\rm C}$\,$=$\,0.0008, $X_{\rm
  O}$\,$=$\,0.0016. In the six panels of Figure~\ref{models} we show
the effects, from top to bottom, of reducing the mass-loss rate to
10$^{-7}$\,$M_\odot$\,yr$^{-1}$, reducing the terminal velocity to
1000\,\kms, introducing a clumped wind with a volume-filling factor,
$f$\,$=$\,0.1 (i.e. 10\%), decreasing the microturblence to 15\,\kms,
reducing the iron abundance by a factor of two, and reducing the
carbon abundance by a factor of ten; changing $\beta$ by $\pm$\,0.2
also had minimal impact. All of the spectra have been convolved with a
rotational broadening profile of 150\,\kms, to match that adopted by
\citet{vfts016}. These preliminary tests suggest that the carbon
abundance is the principal factor influencing the appearance of the
C~\4 line (for this adopted temperature and luminosity), and that our
hypothesis of these stars as carbon-rich (relative to nitrogen) is
plausible\footnote{The PoWR models are not tailored to fit our spectra
  (e.g.  in contrast to the observations, the N~{\scriptsize V} lines
  are predicted in emission for the adopted parameters); our objective
  was a first investigation of the sensitivity of the C~{\scriptsize
    IV} emission to different parameters.}.

\section{Stellar radial velocities}\label{rvs}

To estimate RVs for the O- and B-type stars we used Gaussian fits of
the absorption lines listed in Table~\ref{vrlines}.  The lines
selected are those used by \citet{s13} and \citet{bst} to analyse the
VFTS data, although we chose not to use He~\lam4026 (blend of He~\1
and \2) nor He~\1 \lam4121 (blend with O~\2). We note that He~\1 \lam4471
was also avoided as it is typically the line most affected by nebular
contamination, and also presents other discrepancies for RV estimates,
probably due to its nature as a triplet transition and a possible
blend with O~\2 \citep[for further discussion see Appendix~B
of][]{s13}.

These adopted lines were not available at the earliest and latest
types in the AAOmega sample.  Thus, for the earliest O-type stars
($\le$O4) we used He~\2 \lam\lam4026, 4200, 4542\footnote{At such
  types \lam4026 is strongly dominated by He~{\scriptsize II}
  absorption.}. For the later-type (B9 and A-type) spectra we used
H$\delta$ and H$\gamma$, combined with Si~\2 \lam\lam4128-31 and He~\1
\lam4471.

Mean RVs for each target for the two nights, $v_1$ and $v_2$, are
presented in Table~\ref{30dor}.  The quoted uncertainties are the
standard errors (s.e.) on the means, and the sixth and eighth columns
give the number of measured lines ($n$); mean values are only
calculated for spectra where $n$\,$\ge$\,3. As a check for systematics
in our measurements (e.g. line blends) and the adopted rest
wavelengths, for each line in our analysis we calculated the mean
residuals ($\Delta_{\rm Night1}$ and $\Delta_{\rm Night2}$, and their
standard deviations) compared to the estimated RV for each star; these
are given in the third and fourth columns of Table~\ref{vrlines}. In
general, there are no significant systematic offsets present in the
adopted lines.  The Balmer lines in the cooler spectra appear to yield
(marginally) different RV estimates, but these values are sufficient
for our purposes.

\begin{table}
\begin{center}
  \caption{Rest wavelengths used for the radial-velocity estimates.
    Values in the two final columns are the mean differences ($\Delta$)
    between the velocity estimates from each line and the final
    velocity for each star.}\label{vrlines}
\begin{tabular}{lccc}
\hline\hline
Ion & \lam  & \pp$\Delta_{\rm Night1}$& \pp$\Delta_{\rm Night2}$\\
    & [\AA] & \pp[\kms] & \pp[\kms]\\
\hline
\multicolumn{4}{c}{\em O-type ($\le$\,O4) stars:}\\
 He~\2 & 4025.60  & \pp2.0\,$\pm$\,10.8   & $-$1.5\,$\pm$\,9.3\o \\
 He~\2 & 4199.83  & $-$1.0\,$\pm$\,6.6\o  & \pp3.2\,$\pm$\,6.7\o \\
 He~\2 & 4541.59  & $-$1.0\,$\pm$\,12.2   & $-$1.7\,$\pm$\,4.7\o \\
\hline
\multicolumn{4}{c}{\em O-type ($>$\,O4) stars:}\\
 He~\2 & 4199.83  & $-$1.7\,$\pm$\,12.4  & $-$1.9\,$\pm$\,9.0\o \\
 He~\1 & 4387.93  & \pp0.4\,$\pm$\,9.8\o & \pp0.0\,$\pm$\,8.4\o \\
 He~\2 & 4541.59  & \pp1.8\,$\pm$\,10.4  & $-$0.2\,$\pm$\,7.9\o \\
 He~\2 & 4685.71  & \pp2.2\,$\pm$\,10.7  & \pp2.0\,$\pm$\,10.0 \\
 He~\1 & 4713.15  & $-$1.6\,$\pm$\,9.3\o & \pp\ldots\o \\
 He~\1 & 4921.93  & $-$1.2\,$\pm$\,7.4\o & \pp\ldots\o \\
\hline
\multicolumn{4}{c}{\em B-type stars (excl. B9):}\\
 He~\1 & 4009.26  & \pp0.8\,$\pm$\,8.8\o & \pp1.8\,$\pm$\,9.8\o \\
 He~\1 & 4143.76  & \pp1.0\,$\pm$\,7.1\o & \pp1.1\,$\pm$\,5.7\o \\
 He~\1 & 4387.93  & $-$2.1\,$\pm$\,5.4\o & $-$2.6\,$\pm$\,4.9\o \\
 He~\1 & 4437.55  & $-$0.1\,$\pm$\,8.6\o & \pp0.5\,$\pm$\,9.2\o \\
 Si~\3 & 4552.62  & $-$1.2\,$\pm$\,6.1\o & $-$1.1\,$\pm$\,5.5\o \\
 Si~\3 & 4567.84  & $-$0.7\,$\pm$\,5.2\o & \pp2.1\,$\pm$\,7.7\o \\
 Si~\3 & 4574.76  & $-$0.9\,$\pm$\,7.3\o & $-$1.8\,$\pm$\,7.0\o \\
 He~\1 & 4713.15  & \pp2.4\,$\pm$\,5.4\o & \pp\ldots\o \\
 He~\1 & 4921.93  & \pp0.7\,$\pm$\,6.3\o & \pp\ldots\o \\
\hline
\multicolumn{4}{c}{\em B9/A-type stars:}\\
 H     & 4101.73  & $-$3.6\,$\pm$\,5.2\o & $-$2.4\,$\pm$\,1.8\o \\
 Si~\2 & 4128.07  & \pp3.8\,$\pm$\,11.0  & \pp4.5\,$\pm$\,9.0\o \\
 Si~\2 & 4130.89  & $-$0.3\,$\pm$\,3.2\o & \pp0.9\,$\pm$\,10.9  \\
 H     & 4340.47  & $-$3.7\,$\pm$\,3.3\o & $-$4.2\,$\pm$\,7.6\o \\
 He~\1 & 4471.48  & \pp2.3\,$\pm$\,10.2  & \pp0.3\,$\pm$\,5.0\o \\
\hline
\end{tabular}
\tablefoot{Wavelengths are from the NIST atomic spectra database
  \citep{nist}.}
\end{center}
\end{table}

\subsection{Binaries}\label{binaries}

Our observations were obtained on two (consecutive) nights so our
velocity estimates are somewhat limited in the search for variations
arising from single-lined binaries. Nonetheless, we employed similar
criteria to those used by \citet{s13} and \citet{d15} for
identification of spectroscopic binaries from the VFTS.  For stars
with RV estimates from both nights, we consider it as a spectroscopic
binary if they satisfy that $|v_1$\,$-$\,$v_2|$\,$>$\,20\,\kms\
between the observations and that
$|v_1$\,$-$\,$v_2|$\,$/$\,$\sqrt{\sigma_1^2\,+\,\sigma_2^2}$~$>$~4.0,
i.e. such variations are statistically significant, following the
approach taken by \citet{s13}. The choice of the RV threshold for
binary detection is a trade-off between the number of false positives
arising from the effects of pulsations (and other atmospheric
variations) and detections of real RV shifts from binary motion, as
discussed by \citet{s13} and \citet{d15}.

Employing these criteria, the RV estimates for only one star,
AA$\Omega$~30\,Dor\,053 (classified as B0.5 Ib Nwk), is formally
significant as a SB1 system; this remains the case using a lower
threshold of $|v_1$\,$-$\,$v_2|$\,$>$\,16\,\kms\ \citep[as used by][in
their analysis of early B-type stars]{d15}.  There are a
further six stars\footnote{For completeness:
  AA$\Omega$~30\,Dor\,084, 085, 135, 160, 192, and 419.}  with
$|v_1$\,$-$\,$v_2|$\,$>$\,20\,\kms, but with sufficient uncertainties
on their RV estimates that they do not satisfy the second criterion;
for the purposes of the calculations in the next section we exclude
these potential (though unconfirmed) RV variables.

In summary, eleven SB2 systems were found in our spectroscopy (with an
additional SB2 candidate, AA$\Omega$~30\,Dor\,173), one SB1 system
(AA$\Omega$~30\,Dor\,053), and six potential RV variables; these
objects and their spectral classifications are listed in
Table~\ref{binaries}. Eight are known eclipsing systems from
\citet{g11} from the third phase of the Optical Gravitational Lensing
Experiment (OGLE).  Using a 2$''$ search radius we then cross-matched
our remaining targets with the \citet{g11} catalogue, finding eight
other eclipsing systems that were undetected as binaries from the
available spectra. The OGLE identifiers, periods, and light-curve
classifications are included in Table~\ref{binaries}. Similar
cross-checks with the luminous variables in the LMC reported by
\citet{asasvar} yielded only one match -- AA$\Omega$~30\,Dor\,286
(Sk$-$69$^\circ$238), classified as O7.5 Ib(f), which was detected as
a low-amplitude ($\sim$0.25\,mag) photometric variable.

\begin{table*}
\begin{center}
  \caption{Summary of known binaries and candidate radial-velocity
    variables in the AAOmega sample.}\label{binaries}
\begin{tabular}{lllccc}
\hline\hline
Star & Spectral classification   & AA$\Omega$ status & OGLE LMC-ECL &  P$_{\rm OGLE}$ [d] & Type$_{\rm OGLE}$ \\
\hline
010  & O9.5:~$+$~early B         & SB2            & \ldots       &        \ldots & \ldots \\
021  & B0.5:~$+$~early B         & SB2            & 16629        &  \o\o2.923299 & EC     \\
024  & B1.5 Ib Nwk               & \ldots         & 16646        &   \o26.225620 & ED     \\
027  & B0.5:~$+$~early B         & SB2            & 16675        &  \o\o3.748640 & ED     \\
038  & B1-1.5 V-III              & \ldots         & 16881        &  \o\o3.167171 & ED/VAR \\
053  & B0.5 Ib Nwk               & SB1            & \ldots       &        \ldots & \ldots \\
058  & B1-1.5 V-III              & \ldots         & 17198        &  \o\o3.290830 & ED     \\
065  & O9.2-9.5 V                & \ldots         & 17334        &  \o\o1.362346 & ESD    \\
084  & B1.5 Ib Nwk               & RV var?        & 17823        &  \o\o4.585031 & EC     \\
085  & B0 V                      & RV var?        & \ldots       &        \ldots & \ldots \\
114  & B0.2 III:~$+$~early B     & SB2            & \ldots       &        \ldots & \ldots \\
122  & O9.5 III~$+$~B0:          & SB2            & 18794        &  \o\o5.946846 & ED     \\
130  & B0.2:~$+$~early B         & SB2            & \ldots       &        \ldots & \ldots \\
135  & B1.5 III-II               & RV var?        & \ldots       &        \ldots & \ldots \\
160  & B1.5 V-III                & RV var?        & \ldots       &        \ldots & \ldots \\
173  & B0.5 Ib Nwk?              & SB2?           & 19892        &  \o\o4.682925 & ED     \\
178  & O6 V((f))                 & \ldots         & 19996        &  \o\o1.079433 & ED     \\
192  & O8.5 IIn                  & RV var?        & \ldots       &        \ldots & \ldots \\
267  & O9.5: II~$+$~early B      & SB2            & 20901        &  \o\o1.554470 & ED     \\
330  & O9: V-III~$+$~O9.5: V-III & SB2            & 21568        &  \o\o3.225450 & ED     \\
337  & O9: V-III~$+$~O9.7: V-III & SB2            & \ldots       &        \ldots & \ldots \\
352  & Mid-O V                   & \ldots         & 21844        &    100.374558 & ED     \\
371  & O9.5 II                   & \ldots         & 22166        &  \o\o1.759227 & ED     \\
374  & Early B~$+$~early B       & SB2            & 22270        &  \o\o5.414011 & ED     \\
381  & O9 III                    & \ldots         & 22429        &    100.032930 & ED/VAR \\
414  & O8.5 Iabf ($+$\,OB?)      & SB2            & \ldots       &        \ldots & \ldots \\
419  & B1-1.5 I                  & RV var?        & \ldots       &        \ldots & \ldots \\
\hline
\end{tabular}
\tablefoot{Details of eclipsing binaries known from the OGLE survey
  are from \citet{g11}; classifications of light-curve variability
  in the final column are: EC\,$=$\,contact eclipsing binary;
  ED\,$=$\,detached eclipsing binary; ESD\,$=$\,semi-detached
  eclipsing binary; ED/VAR\,$=$\,detached with additional variability
  superimposed. Classifications of the spectra for
  AA$\Omega$~30\,Dor\,084 and 374 were given by \citet{m14},
  identified in their Table~A3 as lm0020n19615 and lm0031122987,
  respectively; the more detailed classifications given here now
  supercede those.}
\end{center}
\end{table*}

\subsection{Velocity Distributions and Radial-Velocity Outliers}\label{vdist}

We calculated mean velocities of all our targets with RV estimates,
and for subsamples limited to the O- and BA-type spectra. The
resulting means and their associated standard deviations are
summarised in Table~\ref{RV_results}, and the systemic value is in
good agreement with results from the VFTS \citep{bst}. In calculation
of these results we have excluded stars flagged as potential RV
variables in the previous section and those identified as eclipsing
binaries.  We also identified and (by iteration) excluded RV outliers
for both nights, defined as stars with:
$|v\,$-$\,\overline{v}|$\,$>$\,3$\sigma$.

The RV estimates of the nine outliers are summarised in
Table~\ref{rv_out}.  Both AA$\Omega$~30\,Dor\,254 and 383 only have
estimates (from three or more lines) for one night, but qualitative
comparison of the spectra from the two nights reveals no obvious
shifts; indeed, the former of these two objects is VFTS\,016,
identified as a runaway star by \citet[][including analysis of these
data]{vfts016}.  

For completeness, we note that two stars, AA$\Omega$~30\,Dor\,248 and
282, are (marginal) outliers in the estimates from one of the nights,
suggesting either small RV variations or simply that we are at the
limit of the available data (given that relatively few lines were
available for RV estimates for these stars). Following the suggestion
by \citet{w02} that Sk$-$68$^\circ$137 and BI~253 (VFTS\,072,
AA$\Omega$~30\,Dor\,276) might be massive runaways, the case of
AA$\Omega$~30\,Dor\,248 is similarly intriguing given its location
($\sim$22$'$ NNW of R136, see Fig.~\ref{targets}). Indeed, if its
OC-type classification does indicate an early evolutionary phase it
raises the question of whether it is an ejected runaway, or formed
more locally (in relative isolation given the apparent lack of a
nearby star-forming region).

AA$\Omega$~30\,Dor\,159 (W61 28-23) is the second largest outlier, and
we note that \citet{mfast2} reported a comparable RV ($\sim$350\,\kms,
see their Fig. 9) from observations in 1999 January.  Whether this is
a genuine runaway, or just chance observations of similar RVs for a
large-amplitude binary system will require further spectroscopic
monitoring.  Indeed, additional spectroscopy of each object in
Table~\ref{rv_out} will be required to ascertain their true status.

\begin{table}
\begin{center}
{\small
  \caption{Mean radial velocities and dispersions for the AA$\Omega$
    sample for both nights ($\overline{v_1}$ and $\overline{v_2}$,
    respectively), excluding known and candidate binaries (see
    Section~\ref{binaries}) and stars with outlying velocities
    (Table~\ref{rv_out}).}\label{RV_results}
\begin{tabular}{lcccc}
\hline\hline
Sample  & $\overline{v_1}$\,$\pm$\,s.d. [\kms] & $n$ & $\overline{v_2}$\,$\pm$\,s.d. [\kms] & $n$ \\
\hline
All     & 274.1\,$\pm$\,16.4 &  199 & 274.3\,$\pm$\,16.5 &  194  \\
O-type  & 275.1\,$\pm$\,19.5 & \o68 & 275.3\,$\pm$\,18.6 & \o61 \\
BA-type & 273.6\,$\pm$\,14.6 &  131 & 273.8\,$\pm$\,15.5 &  133  \\
\hline
\end{tabular}
}
\end{center}
\end{table}

\begin{table*}
\begin{center}
  \caption{Stars with outlying radial velocities, which are candidate large-amplitude binaries or runaway stars.}\label{rv_out}
\begin{tabular}{llcccccc}
\hline\hline
Star & Spectral type  & $v_1$\,$\pm$\,s.d. & ($v_1$\,$-$\,$\overline{{v_1}}$) & $|v_1$\,$-$\,$\overline{{v_1}}|$/$\sigma$ & 
$v_2$\,$\pm$\,s.d.& ($v_2$\,$-$\,$\overline{{v_2}}$) & $|v_2$\,$-$\,$\overline{{v_2}}|$/$\sigma$ \\
     &                    & [\kms]              & [\kms] &                                                  & [\kms] & [\kms] &             \\
\hline
097  & O9 IIIn           & 209.0\,$\pm$\,15.3  & \o$-$65.1\,$\pm$\,22.4 & 4.0   & 214.7\,$\pm$\,9.5\o & \o$-$59.6\,$\pm$\,19.1 & 3.6   \\
159  & O3.5 III(f$^\ast$)& 370.1\,$\pm$\,5.7\o & \o\pp96.0\,$\pm$\,17.4 & 5.8   & 377.8\,$\pm$\,7.2\o &  \pp103.5\,$\pm$\,18.0 & 6.3   \\ 
195  & B1 Ib Nwk         & 328.3\,$\pm$\,12.2  & \o\pp54.2\,$\pm$\,20.4 & 3.3   & 329.9\,$\pm$\,4.3\o & \pp\o55.6\,$\pm$\,17.1 & 3.4   \\
254  & O2 III-If$^\ast$  & \pp\ldots           & \pp\ldots              & \ldots& 190.3\,$\pm$\,1.6\o & \o$-$84.0\,$\pm$\,16.6 & 5.1   \\
383  & O9.2 Ib           & 189.0\,$\pm$\,10.1  & \o$-$85.1\,$\pm$\,19.3 & 5.2   & \pp\ldots              & \pp\ldots                 & \ldots\\ 
401  & B0.5 Ia           & 222.5\,$\pm$\,7.5\o & \o$-$51.6\,$\pm$\,18.0 & 3.1   & 216.1\,$\pm$\,7.1\o & \o$-$58.2\,$\pm$\,18.0 & 3.5    \\
424  & B0.7 Ib Nwk       & 192.2\,$\pm$\,10.2  & \o$-$81.9\,$\pm$\,19.3 & 5.0   & 199.9\,$\pm$\,7.1\o  & \o$-$74.4\,$\pm$\,18.0 & 4.5    \\
426  & O6.5 III(f)       & 213.6\,$\pm$\,14.5  & \o$-$60.5\,$\pm$\,21.9 & 3.7   & 220.7\,$\pm$\,12.0   & \o$-$53.6\,$\pm$\,20.4 & 3.2    \\
428  & O7.5 V((f))       & 166.4\,$\pm$\,5.8\o & $-$107.7\,$\pm$\,17.4  & 6.6   & 163.7\,$\pm$\,8.1\o  & $-$110.6\,$\pm$\,18.4  & 6.7    \\
\hline
\end{tabular}
\tablefoot{The adopted mean velocities ($\overline{v_1}$ and
  $\overline{v_2}$) are those from the first entry of
  Table~\ref{RV_results} (which exclude the above stars and candidate
  binaries). AA$\Omega$~30\,Dor\,254\,$=$\,VFTS\,016, the
  runaway star reported by \citet{vfts016}.}
\end{center}
\end{table*}

\section{Summary}\label{summary}
We have presented spectral classifications from optical spectroscopy
with 2dF-AAOmega for 263 massive stars in the NE region of the LMC,
together with RV estimates for 233 stars in the sample. Ten stars have
classifications of O4 or earlier, with two (AA$\Omega$\,30\,Dor\,248
and 280) classified as OC-type given the nitrogen deficiency of their
spectra combined with carbon emission; this is the first time such
effects have been seen at such early types.

The spectra of 11 of our targets reveal them as SB2 systems, with a
possible contribution from a secondary component seen in one other
spectrum. From analysis of the RVs estimated from consecutive nights
we identified one SB1 system and six candidate RV variables. Eight of
these 19 objects are known eclipsing binaries from the OGLE survey
\citep{g11}. Eight of our other targets were also classified as
eclipsing systems by \citet{g11}, but were undetected as binaries from
our spectroscopy.

Using a 3$\sigma$ threshold compared to the systemic velocity (and
excluding the known and candidate binaries), the estimated RVs were
used to identify nine RV outliers (Table~\ref{rv_out}). These are
likely to be large-amplitude binaries or runaway stars, and follow-up
spectroscopy is required to clarify their nature.

\begin{acknowledgements}
  We are grateful to Nolan Walborn for his careful reading of the
  draft manuscript, and for his thoughts regarding the classification
  of the early-type spectra. We also thank the referee for their
  constructive comments, and Russell Cannon and Gary Da~Costa for
  their help with the observations.
\end{acknowledgements}

\bibliographystyle{aa}
\tiny
\bibliography{25882}

\end{document}